\documentclass[modern]{aastex63}

\usepackage{amsmath}

\usepackage[nonumberlist]{glossaries}
\setacronymstyle{long-short}

\newacronym{rte}{RTE}{radiation transfer equations}
\newacronym{see}{SEE}{statistical equilibrium equations}
\newacronym{los}{LOS}{line of sight}
\newacronym{lc}{LC}{long characteristics}
\newacronym{sc}{SC}{short characteristics}
\newacronym{clv}{CLV}{center-to-limb variation}
\newacronym{3d}{3D}{three-dimensional}

\newcommand\figa{Fig.~}

\received{January 26, 2021}
\revised{February 19, 2021}
\accepted{March 5, 2021}

\submitjournal{ApJ}

\shorttitle{Long characteristics}
\shortauthors{de Vicente et al.}

\begin{document}

\title{Long characteristics vs. short characteristics \\ in 3D radiative
  transfer simulations of polarized radiation}

\correspondingauthor{Ángel de Vicente}
\email{angel.de.vicente@iac.es}

\author{A. de Vicente}
\affiliation{Instituto de Astrofísica de Canarias, E-38200 La Laguna, Tenerife, Spain}
\affiliation{Universidad de La Laguna, Departamento de Astrofísica, E-38206, La Laguna, Tenerife, Spain}
  
\author{T. del Pino Alemán}
\affiliation{Instituto de Astrofísica de Canarias, E-38200 La Laguna, Tenerife, Spain}
\affiliation{Universidad de La Laguna, Departamento de Astrofísica, E-38206, La Laguna, Tenerife, Spain}

\author{J. Trujillo Bueno}
\affiliation{Instituto de Astrofísica de Canarias, E-38200 La Laguna, Tenerife, Spain}
\affiliation{Universidad de La Laguna, Departamento de Astrofísica, E-38206, La Laguna, Tenerife, Spain}
\affiliation{Consejo Superior de Investigaciones Científicas, Spain}

\begin{abstract}
We compare maps of scattering polarization signals obtained from
three-dimensional (3D) radiation transfer calculations in a magneto-convection
model of the solar atmosphere using formal solvers based on the ``short
characteristics'' (SC) and the ``long characteristics'' (LC) methods. The SC
method requires less computational work, but it is known to introduce spatial
blurring in the emergent radiation for inclined lines of sight. For polarized
radiation this effect is generally more severe due to it being a signed quantity
and to the sensitivity of the scattering polarization to the model's
inhomogeneities. We study the differences in the polarization signals of the
emergent spectral line radiation calculated with such formal solvers. We take as
a case study already published results of the scattering polarization in the
\ion{Sr}{1} 4607~\AA\ line obtained with the SC method, demonstrating that in
high-resolution grids it is accurate enough for that type of study. In general,
the LC method is the preferred one for accurate calculations of the emergent
radiation, reason why it is now one of the options in the public version of the
3D radiative transfer code PORTA.
\end{abstract}


\section{Introduction}\label{sec:intro}

The modeling of the scattering polarization in spectral lines using \gls*{3d}
models of stellar atmospheres requires the solution of the radiative transfer
problem without assuming local thermodynamic equilibrium. The \gls*{see} and the
\gls*{rte} must be solved iteratively, as both sets of equations are coupled and
the problem as a whole is non-linear and non-local.

In practice, the solution of the \gls*{rte} is the most computationally
demanding part.  Nowadays, the \gls*{rte} are usually solved via the \gls*{sc}
method (for a review of the methods used prior to the \gls*{sc} method see,
e.g.,
\citealt{Jones1973,JonesSkumanich1973,CrosbieLinsenbardt1978}). \cite{Kunasz1988}
developed the \gls*{sc} method based on the parabolic approximation of the
source function. Different approximations of the source function that avoid the
introduction of new local extrema were later introduced, such as piecewise
cubic Hermite interpolation and monotonic Bezier splines
(\citealt{Auer2003,Stepan2013,DelaCruz2013}).

The main advantage of the \gls*{sc} method is its efficiency, as it scales with
just the number of grid points. However, this method is known to introduce
diffusion (\citealt{Kunasz1988,Leenaarts2020}). While in practice this diffusion
is not a problem regarding the calculation of the mean radiation field within
the \gls*{3d} model, this is not necessarily the case when calculating the
linear polarization caused by the scattering of anisotropic radiation. Not only
is the polarization a signed quantity (thus prone to signal cancellation), but
it depends dramatically on the geometry and the spatial variation of the plasma
properties (e.g., \citealt{Manso2011}).

In this paper we study the effect of the diffusion in the \gls*{sc} method in
the formal solution to compute the emergent Stokes profiles in a \gls*{3d} model
of the solar atmosphere. In \S\ref{sec:charmethods} we briefly describe the
\gls*{sc} and \gls*{lc} methods and we give some details on the implementation
of such methods in the radiation transfer code PORTA. The comparison of the
results with the two methods is shown in \S\ref{sec:comparison-sc-lc}. Finally,
we present our conclusions in \S\ref{sec:conclusions}.

\section{The SC and LC methods}\label{sec:charmethods}

The \gls*{rte} for the Stokes-$I$ parameter, which describes the propagation of
the intensity of a radiation beam through a medium for a given frequency and
direction, can be written as (e.g., \citealt{Mihalasetal1978})

\begin{equation}
  \frac{dI}{ds} = -\eta_II + \epsilon_I,
\label{RTE}
\end{equation}
where $I$ is the intensity Stokes parameter, $\eta_I$ is the extinction
coefficient and $\epsilon_I$ is the emissivity. Introducing the optical depth
$d\tau = -\eta_I ds$, the intensity of the beam of radiation of a given
frequency after travelling through a medium an optical depth $\Delta\tau$ in a
given propagation direction is

\begin{equation}
I(\Delta\tau) = I(0)e^{-\Delta\tau} + \int^{\Delta\tau}_{0}S e^{-t}dt,
\label{RTESol}    
\end{equation}
where $I(0)$ is the intensity at the beginning of the propagation path and
$S=\epsilon_I/\eta_I$ is the source function. Equations
\eqref{RTE}-\eqref{RTESol} are easily generalized to the equations for the four
Stokes parameters $I$, $Q$, $U$, and $V$ we actually use (see
\citealt{Landi2004}). However, the intensity only equations are enough for the
purpose of this section.

The ``characteristics'' methods solve Eq. \eqref{RTESol} by assuming a
functional form for the source function $S$. However, the \gls*{sc} and
\gls*{lc} methods differ on whether we use short or long characteristics for
each ray direction, as explained below.

In the \gls*{sc} method, for every point of the grid we take the closest
intersections between the propagating beam and the surfaces between the
surrounding grid points, both forward and backward along the propagation
direction. Because, in general, only the central point under consideration
corresponds exactly with a grid point, all quantities in Eq. \eqref{RTESol}
($\eta_I$, $\epsilon_I$, and $I(0)$) must be interpolated from the surrounding
grid points into the intersections \citep{Kunasz1988}.

As for the \gls*{lc} method, for every point of the grid the beam is propagated
backwards until the boundary of the computational domain is reached. Then,
Eq. \eqref{RTESol} is sequentially solved along the ray. Generally, as in the
\gls*{sc} method, only the point under consideration coincides with a grid point
and, therefore, interpolation is necessary. However, while both \gls*{sc} and
\gls*{lc} require the interpolation of the radiation transfer coefficients
$\eta_I$ and $\epsilon_I$ at every surface between the model's grid points, the
\gls*{lc} method only requires the interpolation of $I(0)$ at the boundary of
the domain.

\begin{figure}
\epsscale{0.8}
\plotone{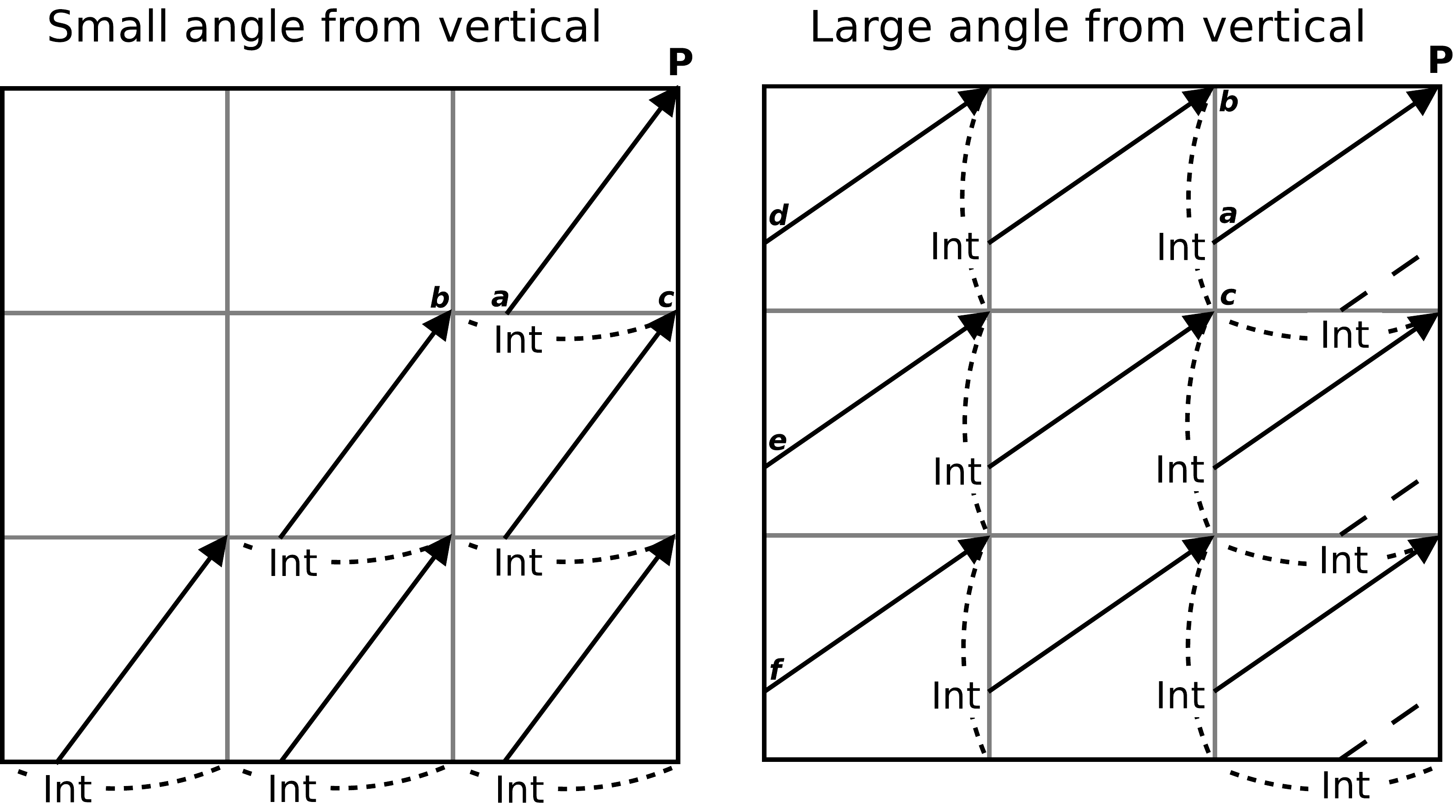}
\caption{Diffusion increases for large angles from the vertical with the
  \gls*{sc} method. The ``Int'' labels in the figure indicate where it is
  necessary to interpolate the intensity ($I(0)$ in Eq. \eqref{RTESol}). See
  full explanation in main text.}
\label{fig:fig_sc_inclined}
\end{figure}

Given a cubic domain of side $N$, the computational work of the \gls*{sc} method
scales as $N{}^3$, while that of the \gls*{lc} method scales as $N{}^4$. The
\gls*{sc} method is significantly computationally less expensive than the
\gls*{lc} one, but in coarse grids it introduces numerical diffusion due to the
high number of upwind interpolations of the intensity needed. This is even more
acute when the propagation direction has a large angle from the vertical, in
which case it cuts vertical planes, for which further interpolations of the
intensity ($I(0)$ in Eq. \eqref{RTESol}) are needed, which in turn increase the
diffusion. In \figa \ref{fig:fig_sc_inclined} we show the interpolations required in order
to calculate the intensity at a given grid point $\mathbf{P}$ for propagation
directions with a small (left panel) and a large (right panel) angle with respect
to the vertical, assuming periodic horizontal boundary conditions. The ``Int''
labels in the figure indicate where it is necessary to interpolate the incoming
intensity ($I(0)$ in Eq. \eqref{RTESol}), and the dotted curves indicate the
points needed for the interpolation (assuming linear order interpolations).
For propagation directions close to the vertical, calculating the intensity at
grid point $\mathbf{P}$ requires the intensity at point $a$ in the figure,
which needs to be interpolated from the values at the grid points $b$ and $c$.
These points, in turn, need the intensity values from the horizontal plane
immediately below, which also need to be interpolated from their closest grid
points. This procedure is propagated backwards until we reach the bottom
plane, where the given boundary condition is used in the interpolations.
Therefore, the calculation of the intensity at point $\mathbf{P}$ as depicted
in \figa \ref{fig:fig_sc_inclined} would require six interpolations.
Likewise, for a significantly inclined propagation direction, the calculation of
the intensity at grid point $\mathbf{P}$ requires the intensity at point $a$ in the
figure, and this value needs to be interpolated from the intensity values at the
$b$ and $c$ grid points. The procedure is analogous to the previous case
until we reach the left vertical boundary plane. Points $d$, $e$, and $f$ in the
figure cannot be interpolated from boundary condition values. Instead, these
rays propagate cycling through the domain (dashed lines in the figure). At the
next intersection with a vertical or horizontal plane (as in the figure), further
interpolations are required. Thus, the calculation of the intensity for the
large angle case at point $\mathbf{P}$ as depicted in the figure would require
nine interpolations.

\begin{figure}
\plotone{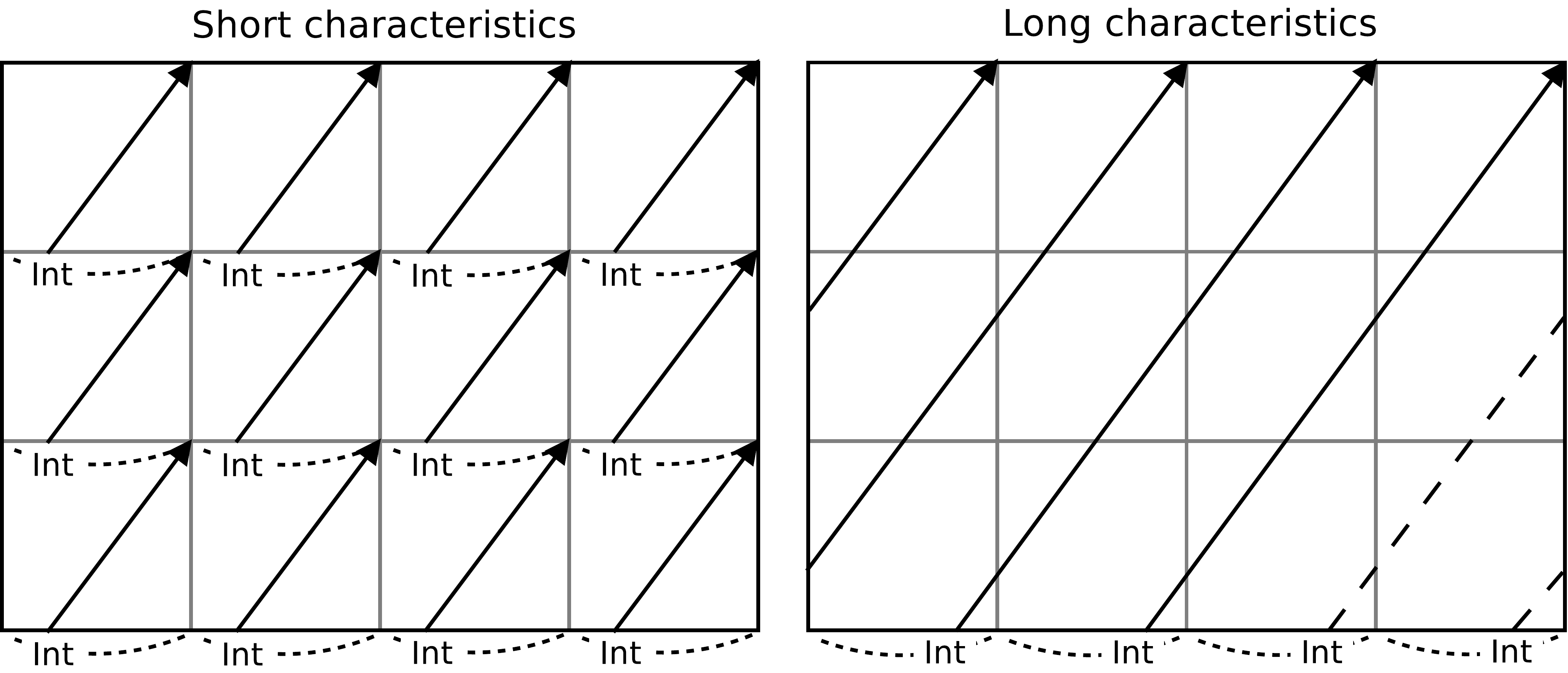}
\caption{Propagation strategy to compute the radiation emerging at the top
  plane with the \gls*{sc} and the \gls*{lc} methods. The ``Int'' labels in the figure indicate
  where it is necessary to interpolate the intensity ($I(0)$ in
  Eq. \eqref{RTESol}). The interpretation of the dotted curves and the dashed
  lines are as per \figa \ref{fig:fig_sc_inclined}.}
\label{fig:fig_sc_vs_lc_rays}
\end{figure}

\figa \ref{fig:fig_sc_vs_lc_rays} shows the difference between the \gls*{sc} and
\gls*{lc} methods in 2D. With the \gls*{sc} method (left panel), we need to interpolate
the intensity values for each short ray. With the \gls*{lc} method (right panel), we only
need to interpolate the intensity values in the bottom boundary, which are then
propagated through the domain (as in \figa \ref{fig:fig_sc_inclined} the dashed
lines represent the continuation of rays that leave the left boundary).

Depending on the particular sampling of the model and the inclination of the
propagation rays, the difference between the solution calculated with \gls*{sc}
and \gls*{lc} can be quite substantial, as we will show in following sections.

\subsection{PORTA}\label{sec:porta}

PORTA is a 3D radiation transfer code capable of calculating the intensity and
polarization of the emergent radiation taking into account scattering processes
and the Hanle and Zeeman effects (\citealt{Stepan2013}, publicly available at
\url{https://gitlab.com/polmag/PORTA}).

\begin{figure*}
\gridline{\fig{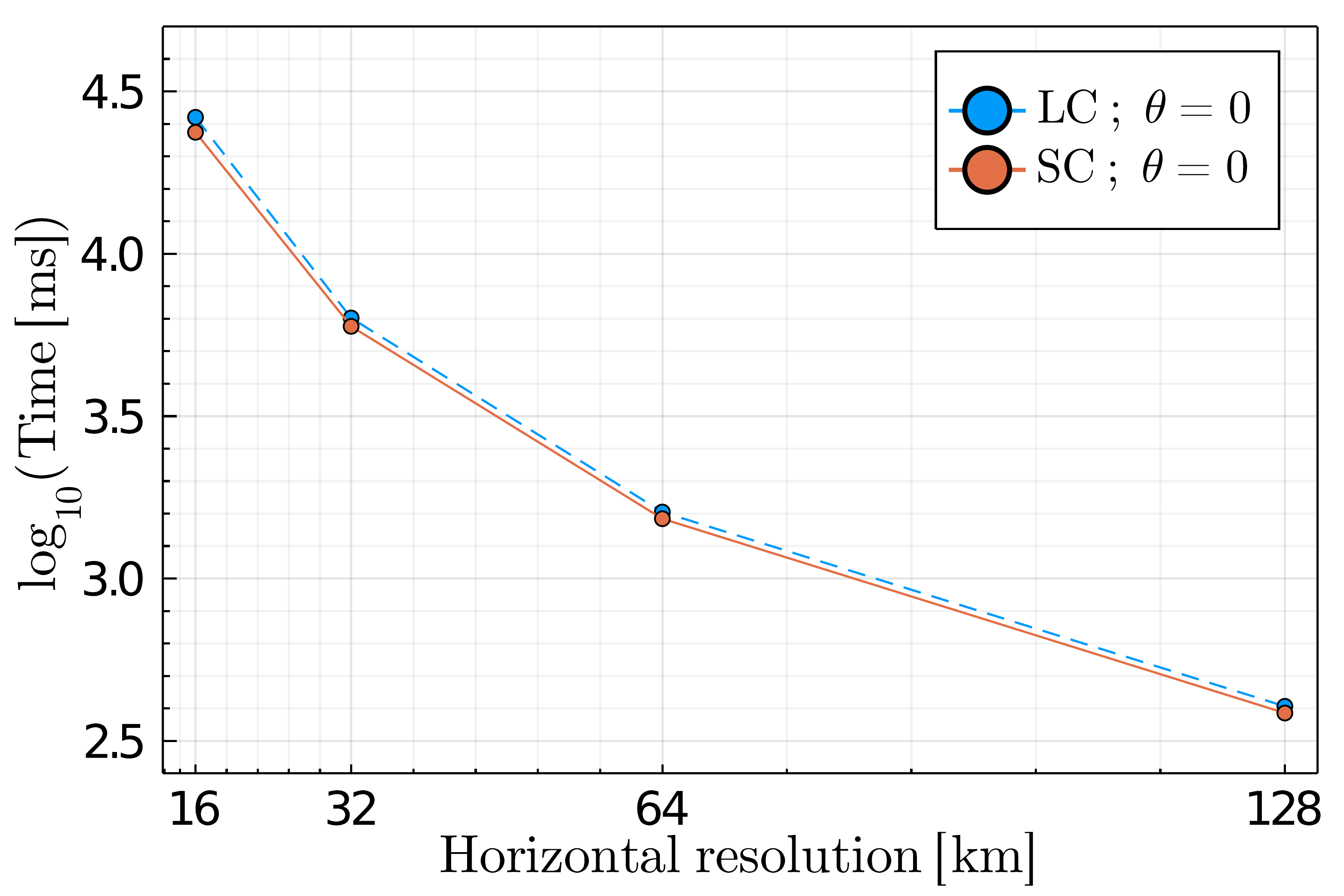}{0.48\textwidth}{(a)}
          \fig{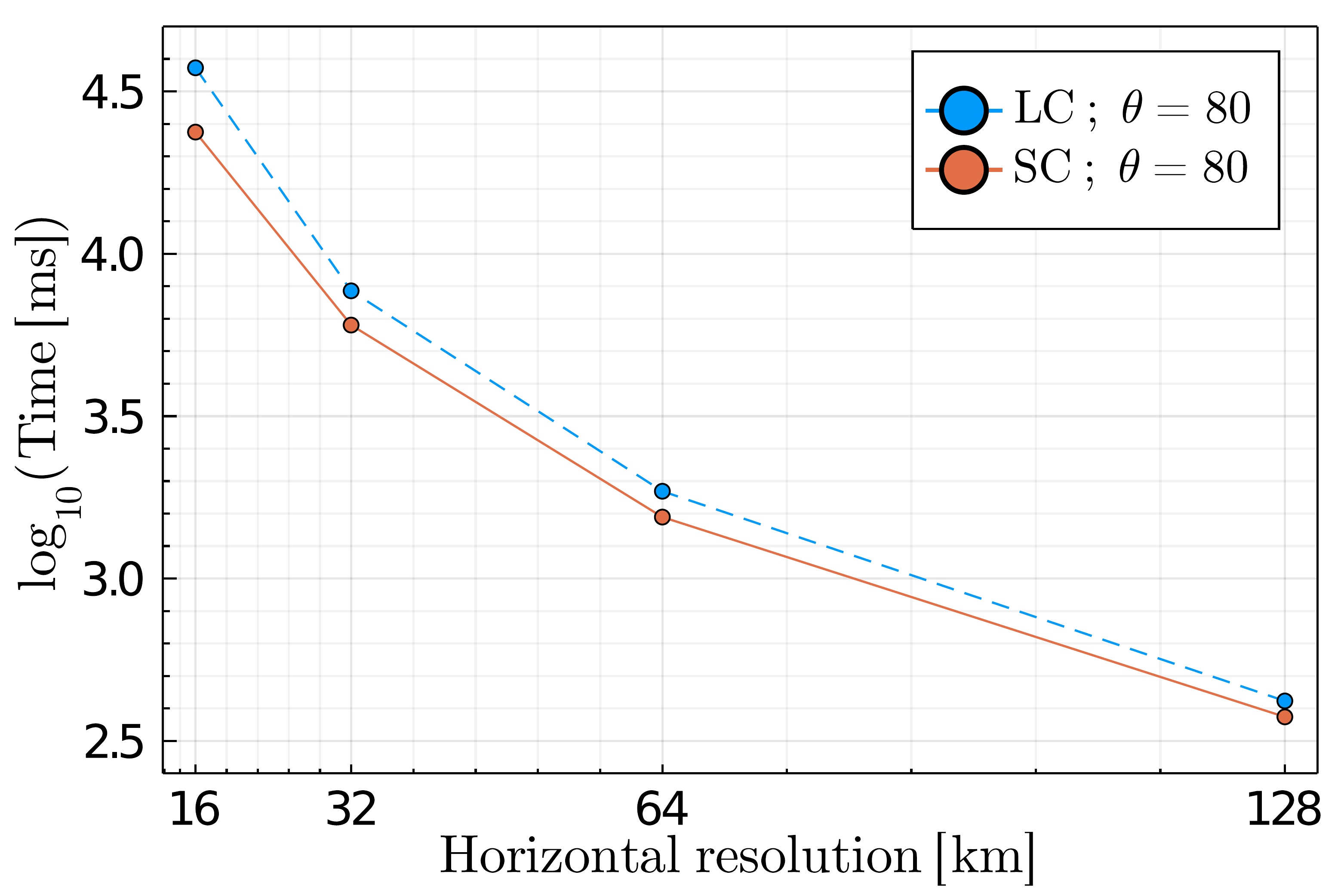}{0.48\textwidth}{(b)}
          }
\caption{Computing time required with the PORTA code for one formal solution of
         the Stokes parameters in the \ion{Sr}{1} 4607~\AA\ line in model
         atmospheres with four different horizontal spatial
         resolutions, ranging from 16~km to 128~km, and constant vertical
         resolution of 16~km. The time difference is small for small inclinations
         $\theta$ with respect to the vertical and increases with it. For
         (a) $\theta=0$ the maximum relative difference between computing times
         is $\sim11$\%, while for (b) $\theta=80$ it is $\sim58$\%.}
\label{fig:fig_ctime_lc_vs_sc}
\end{figure*}

This numerical code solves the radiation transfer problem iteratively with the
Jacobi method. In order to compute the radiation field at every point,
frequency, and direction within the atmospheric model, PORTA solves the
\gls*{rte} with the \gls*{sc} method. Once the self-consistent solution has been
achieved, we can obtain the emergent Stokes profiles at every point of the
model's surface for the desired \gls*{los} with a single formal solution , which
can be efficiently performed with either the \gls*{sc} or the \gls*{lc}
method. The computing time penalty incurred by using the \gls*{lc} instead of
the \gls*{sc} method for a single formal solution is larger as the inclination
$\theta$ with respect to the vertical increases, but never by more than
$\sim58$\% for the performed tests (see \figa \ref{fig:fig_ctime_lc_vs_sc}, for
the \gls*{3d} model described in \S\ref{sec:comparison-sc-lc}).  Note that the
\gls*{lc} method is never used in PORTA when looking for the self-consistent
solution because this would require, for every iterative step, to compute the
radiation field not only at the model's surface, but for all grid points, which
would make the computing time required for solving the problem completely
prohibitive.

Previous results using the PORTA code, such as those by
\citet{DelPinoAleman2018}, made use of the \gls*{sc} method to compute the
emergent Stokes profiles. Recently, the \gls*{lc} method for the formal solution
was implemented in the public version of PORTA, making it possible to test the
accuracy of the results obtained using the faster \gls*{sc} method.

\subsection{3D Test of a Beam Propagating in Vacuum}\label{sec:vacuum}

One of the simplest tests to demonstrate the numerical dispersion of the
radiation is the propagation of a single beam in vacuum (\citealt{Auer1994}). We
have performed this test with both the \gls*{sc} and \gls*{lc} methods using
several grids with different box sizes and resolutions. The
extinction and source functions are set to zero everywhere
(vacuum), as well as the incoming radiation in the boundaries, except for a
single point in the centre of the bottom boundary, and only in one propagation
direction.

When the propagation direction is along the vertical, all points along the
characteristics correspond to nodes in the model. Therefore, there is no
dispersion and \gls*{sc} and \gls*{lc} are equivalent. However, when the
propagation direction is not parallel to any of the axis of the Cartesian grid,
interpolations are needed.

\begin{figure*}
  \gridline{\fig{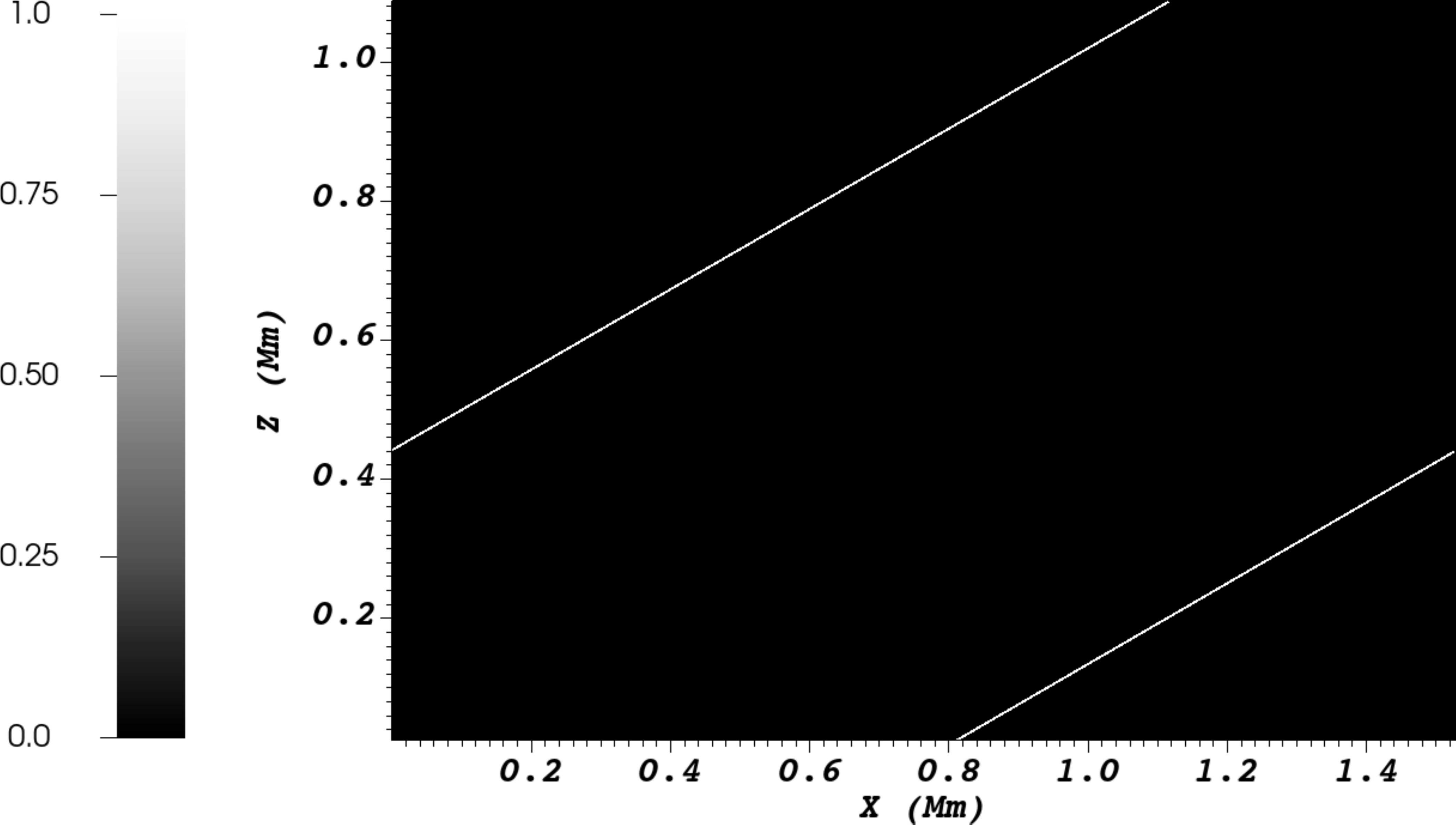}{0.48\textwidth}{(a)}
    \fig{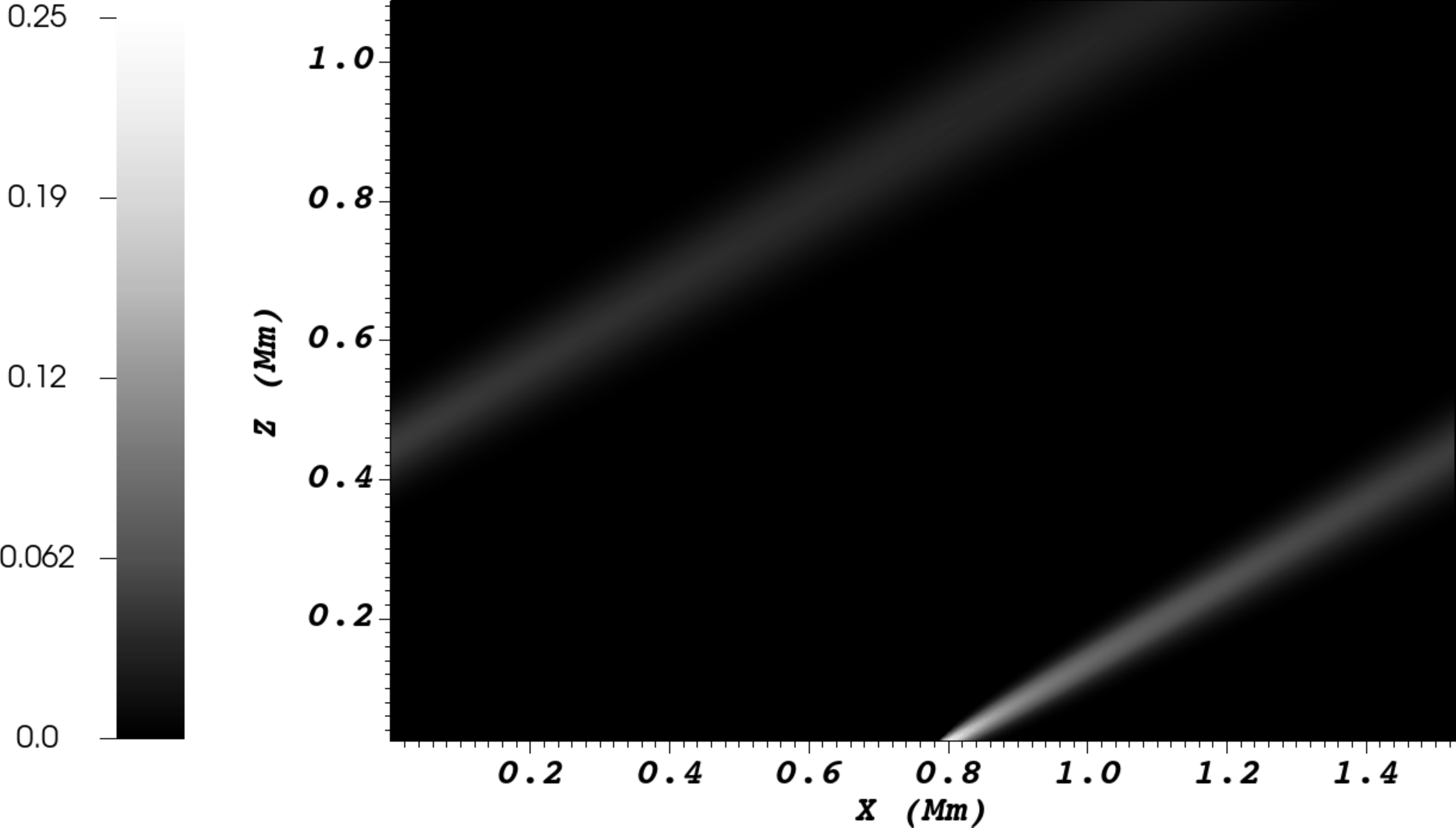}{0.48\textwidth}{(b)}}
  \gridline{\fig{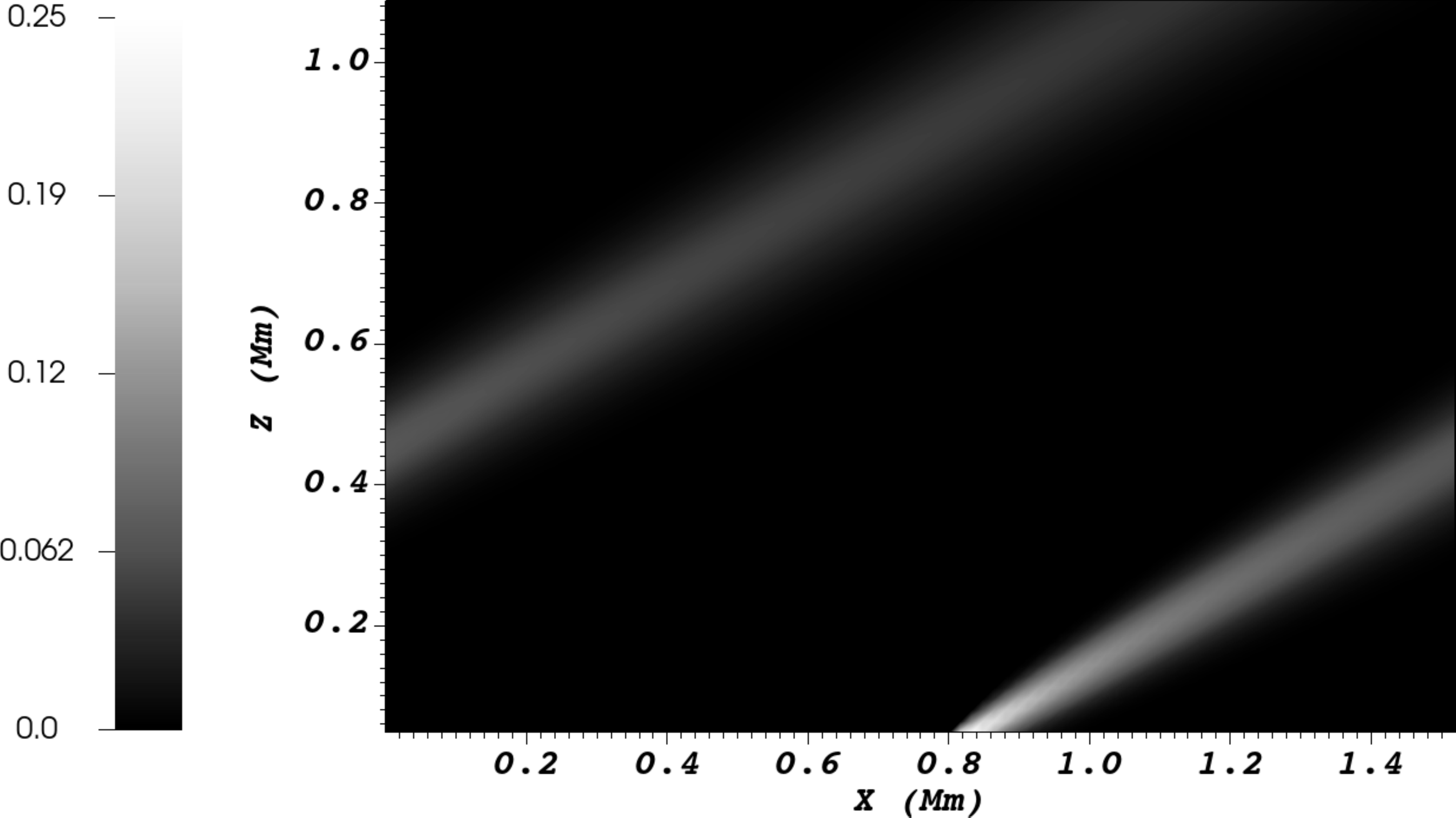}{0.48\textwidth}{(c)}
    \fig{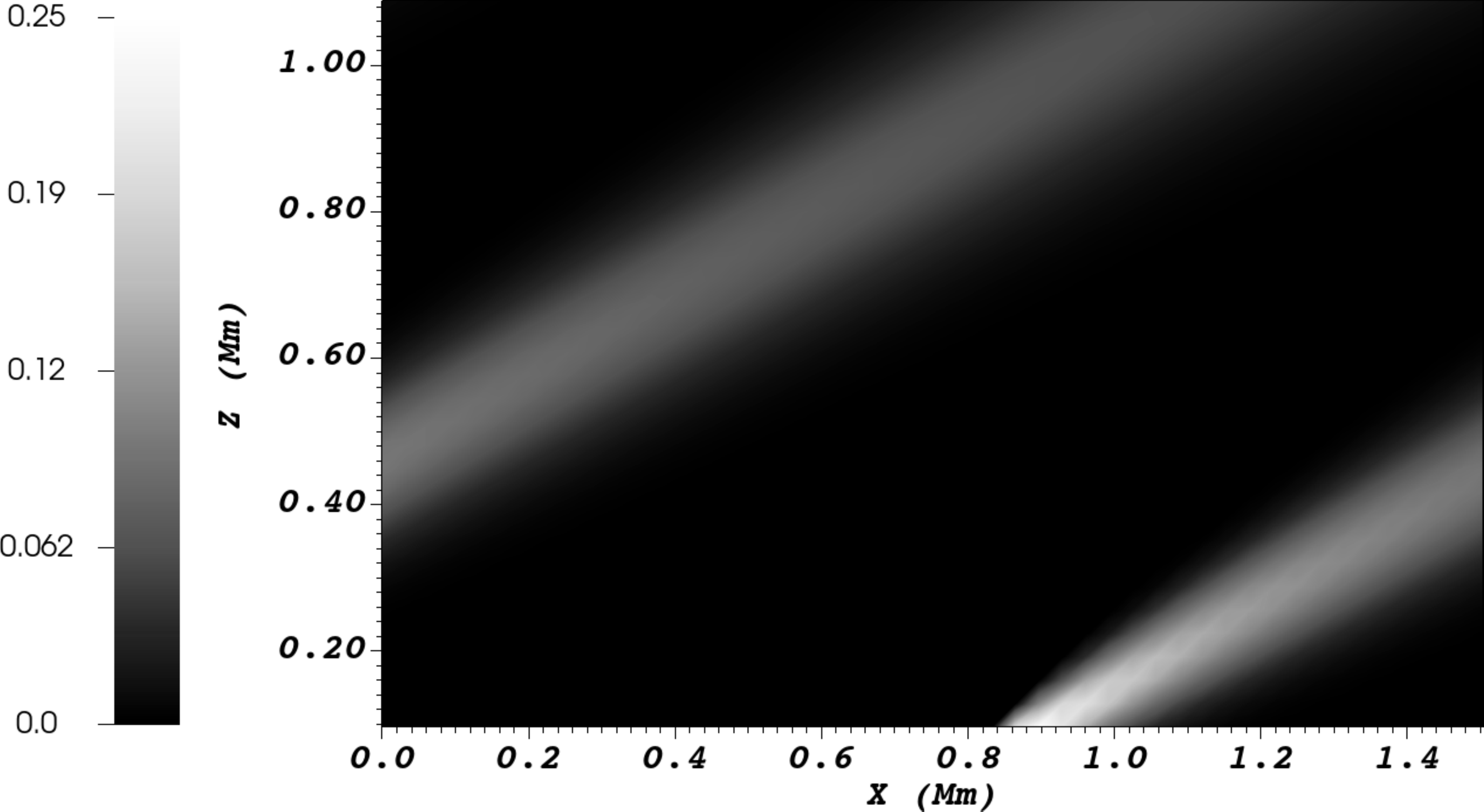}{0.48\textwidth}{(d)}}
  \caption{Propagation of a beam of radiation in vacuum along a direction
    inclined $60^\circ$ from the vertical axis in a Cartesian grid computed
    with: (a) \gls*{lc}; (b), (c) and (d) \gls*{sc}. The domain is
    $\sim1.5\times1.5\times1$~Mm${}^3$ in all panels. The \gls*{lc} solutions are
    virtually identical for all grid resolutions, since the only intensity
    interpolation happens at the bottom boundary, so we only show the case for
    resolution of $8\times8\times8$~km (a). In the case of \gls*{sc}, the
    dispersion is different depending on the grid resolution and inclination of
    the beam, and we present three cases: $8\times8\times8$~km (b),
    $16\times16\times16$~km (c) and $32\times32\times32$~km (d), all saturated
    at 25\% of the original beam intensity for ease of comparison.}
\label{fig:fig_inclined_ray_periodic_lc_vs_sc}
\end{figure*}

\figa \ref{fig:fig_inclined_ray_periodic_lc_vs_sc} shows the case of a
propagation direction inclined $60^\circ$ from the vertical, using a domain
size of $\sim1.5\times1.5\times1$~Mm${}^3$.
Even when the grid has very high resolution ($8$~km), the dispersion in the
calculation with \gls*{sc} is evident (see top row in the figure).  This
dispersion is stronger when the length of the propagation path increases and
when the grid resolution decreases (see bottom row in the figure for the $16$~km
and $32$~km resolution grids).

\begin{figure*}
\gridline{\fig{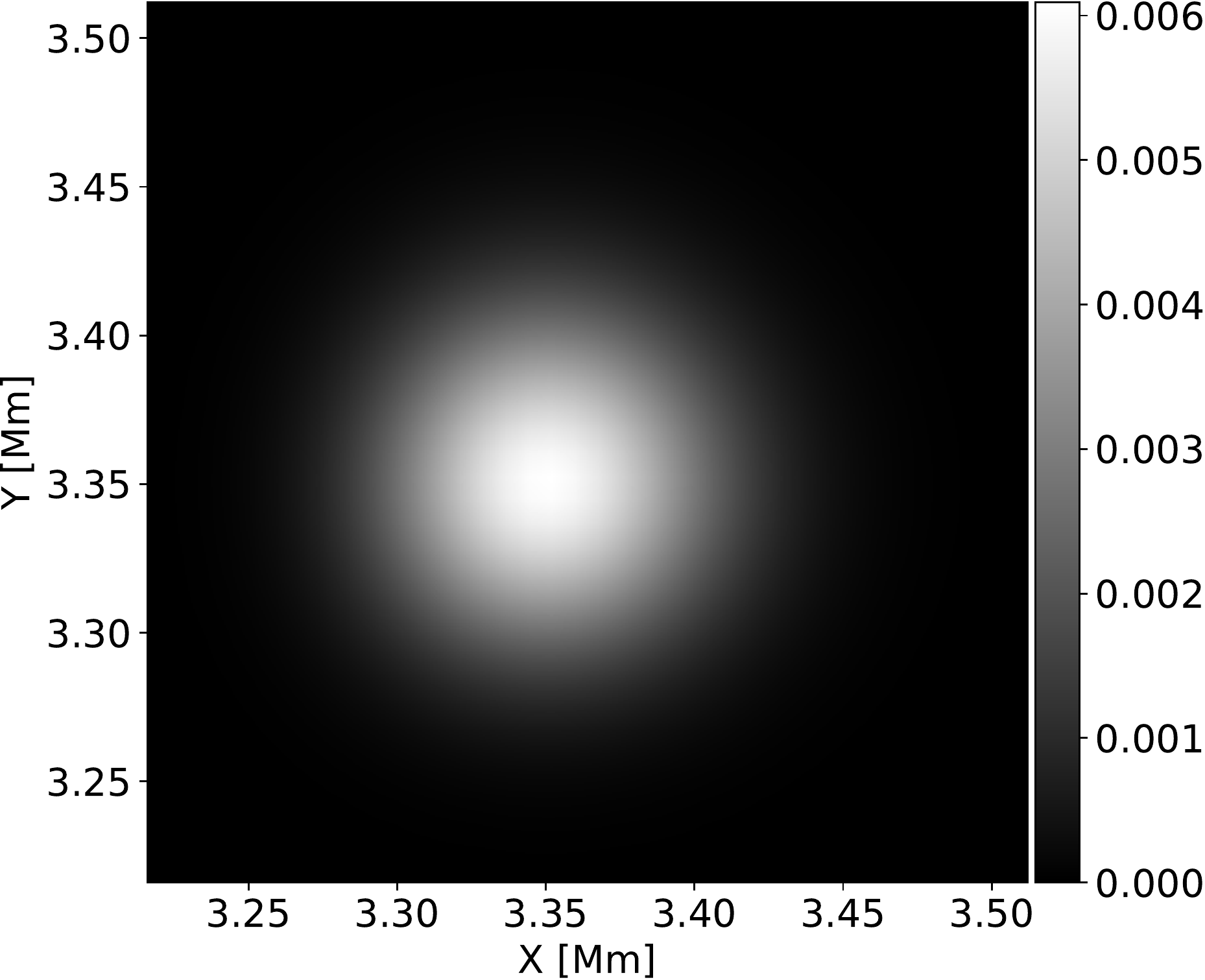}{0.48\textwidth}{(a)}
  \fig{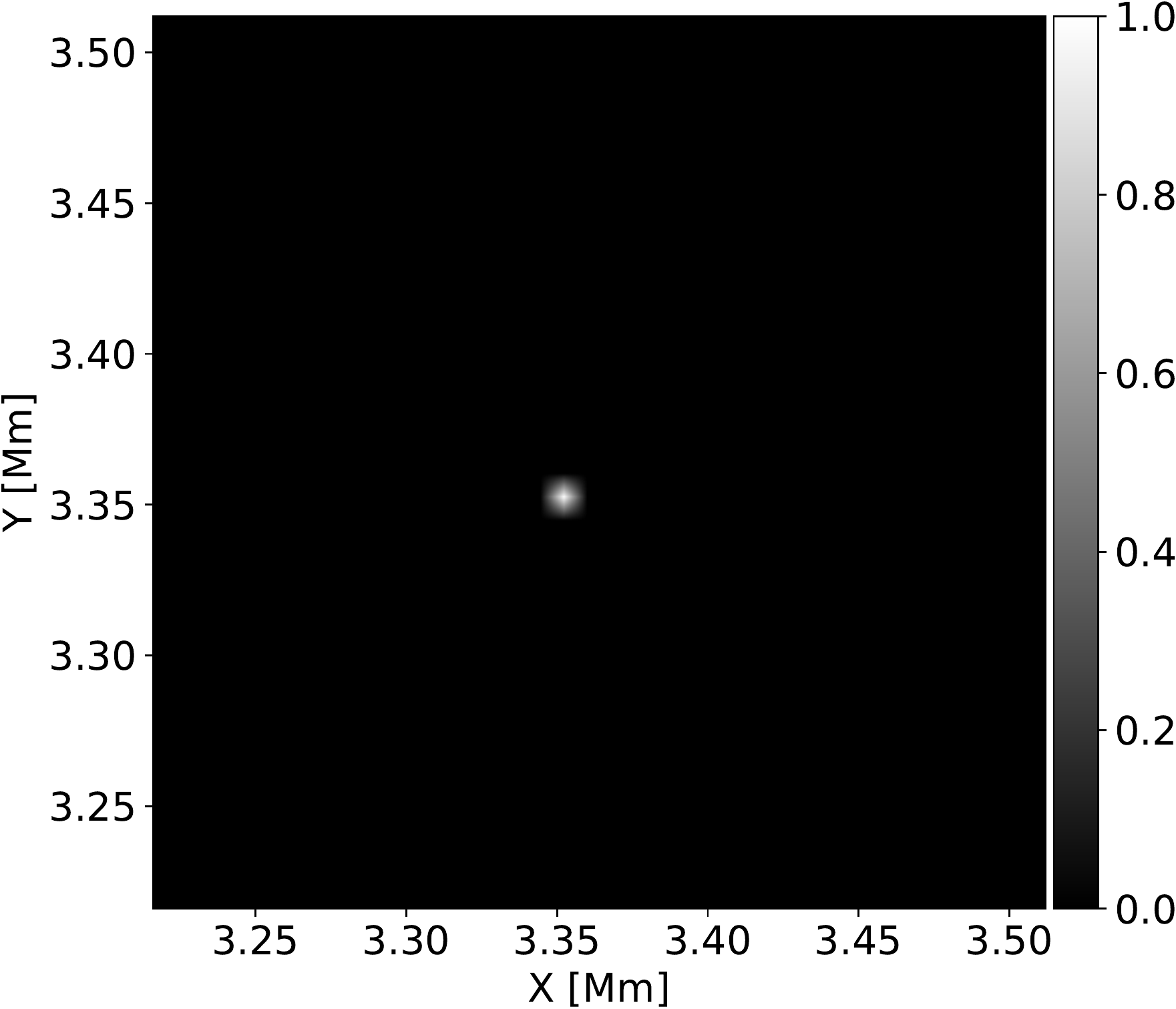}{0.46\textwidth}{(b)}}
\caption{Emergent intensity at the top boundary for a beam of radiation of $I=1$
  in vacuum propagating in a direction inclined $20^\circ$ from the vertical
  axis in a Cartesian grid of $8\times8\times8$~km resolution using (a)
  \gls*{sc} and (b) \gls*{lc}.}
\label{fig:fig_inclined_ray_lc_vs_sc}
\end{figure*}

We have also performed this test with the same grid as the one used in
\citet{DelPinoAleman2018} (a $\sim6\times6\times1$~Mm${}^3$ box with $8$~km
resolution in the three dimensions), and \figa
\ref{fig:fig_inclined_ray_lc_vs_sc} shows the emergent intensity at the top
boundary of the corresponding vacuum box. It is clear from this figure that the
emergent Stokes parameters will suffer from ``smearing'' effects whenever
\gls*{sc} is used to compute the formal solution, while the \gls*{lc} algorithm
prevents such effects.

\section{Comparing the LC and SC results}\label{sec:comparison-sc-lc}

We take as a test case the \ion{Sr}{1} 4607~\AA\ line investigation by
\citet{DelPinoAleman2018} where the \gls*{sc} formal solver was used both in the
iterative solution and in the calculation of the Stokes profiles of the emergent
radiation after obtaining the self-consistent solution of the radiation transfer
problem. To confront the \gls*{lc} and \gls*{sc} methods, we compare the ensuing
scattering polarization signals of the emergent spectral line radiation
calculated in a \gls*{3d} magneto-hydrodynamical model of the quiet solar
photosphere by \citet{Rempel2014}. The model is characterized by a mean field
strength $\langle B
\rangle\approx 170$~G at the model's visible surface,
     with a physical size of $\sim6\times6\times1$~Mm${}^3$ (in a grid of $768\times768\times137$ points
     and $8$~km of spatial resolution).
Further details about this model of the solar photosphere can be found in
\citet{Rempel2014} and \citet{DelPinoAleman2018}. Starting from
  the self-consistent solution calculated by \citet{DelPinoAleman2018} in the
  \gls*{3d} model of \citet{Rempel2014}, we calculate the emergent Stokes
  parameters using the \gls*{lc} formal solver. We then compare these results with the ones shown in
\citet{DelPinoAleman2018}, computed using the \gls*{sc} formal solver.
We used a two-level atomic model with the \ion{Sr}{1} ground level ${}^1$S${}_{0}$ and the
line’s upper level ${}^1$P${}_{1}$, whose angular momentum values are $J_l=0$ and $J_u=1$,
respectively, and with total number densities calculated with a more
complex model atom (see \citet{DelPinoAleman2018} for details).

\subsection{Impact on the center-to-limb variation (CLV)}

\begin{figure}
  \plotone{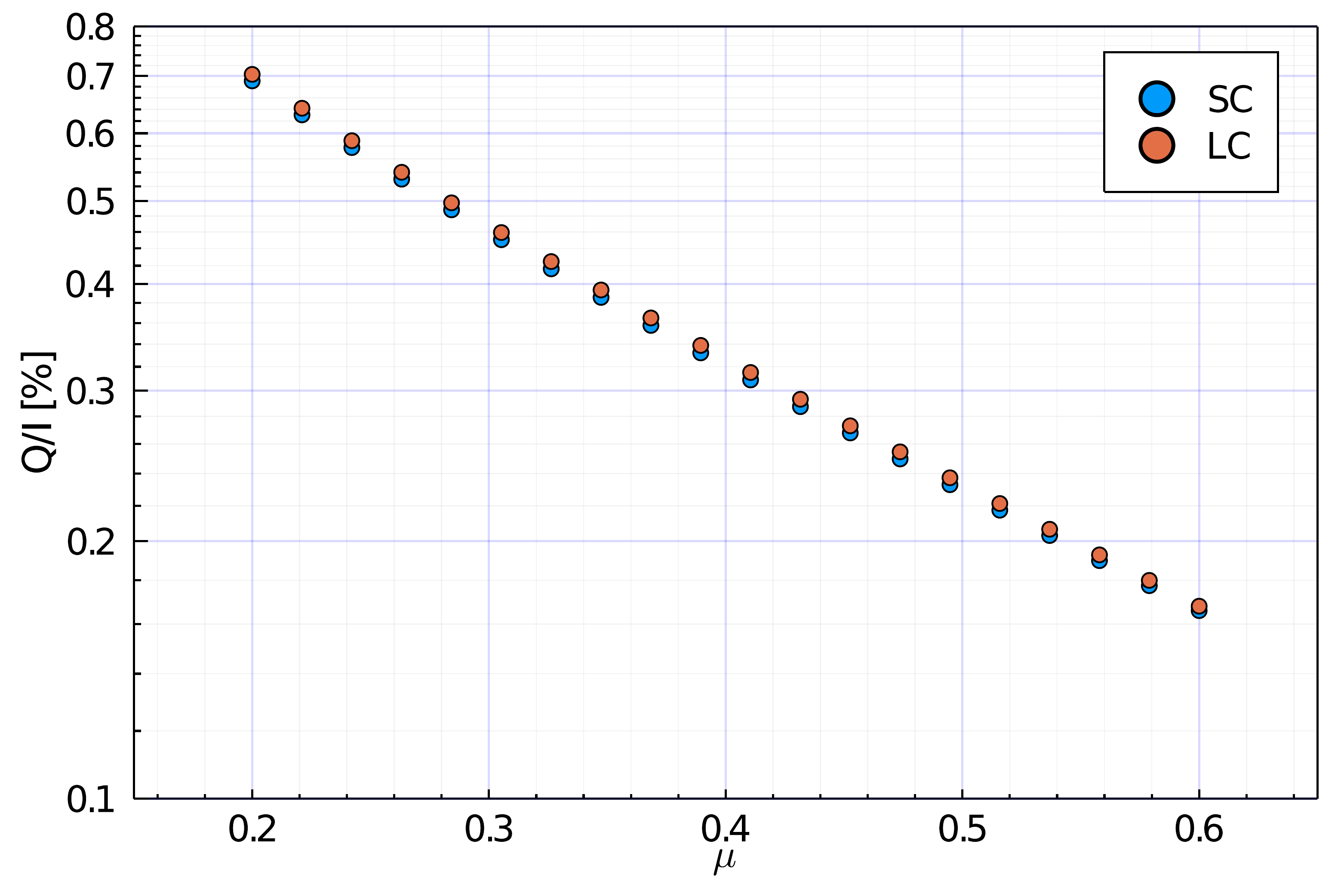}
  \caption{The \gls*{clv} of the $Q/I$ fractional polarization amplitude
  calculated in the $8$~km grid after spatially averaging the Stokes $I$ and $Q$ signals, 
  using the \gls*{sc} method (blue dots) and the \gls*{lc} method (orange
  dots). Each data point in the figure was obtained by integrating over
    twelve equally spaced azimuths and over the whole field of view
    corresponding to each line of sight characterized by $\mu$.}
  \label{fig:fig_clv_lc_vs_sc}
\end{figure}

\begin{figure*}
\gridline{\fig{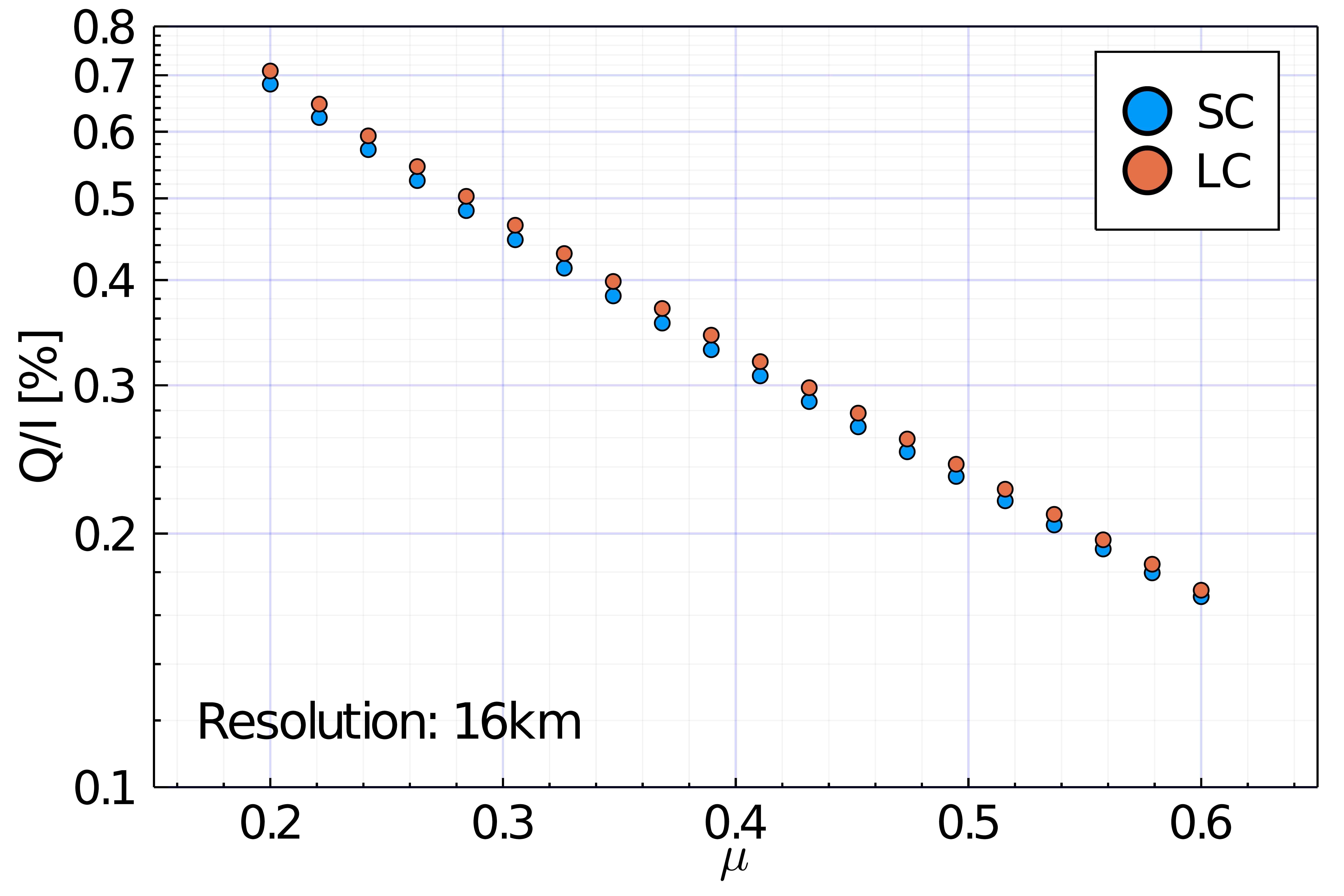}{0.48\textwidth}{(a)}
          \fig{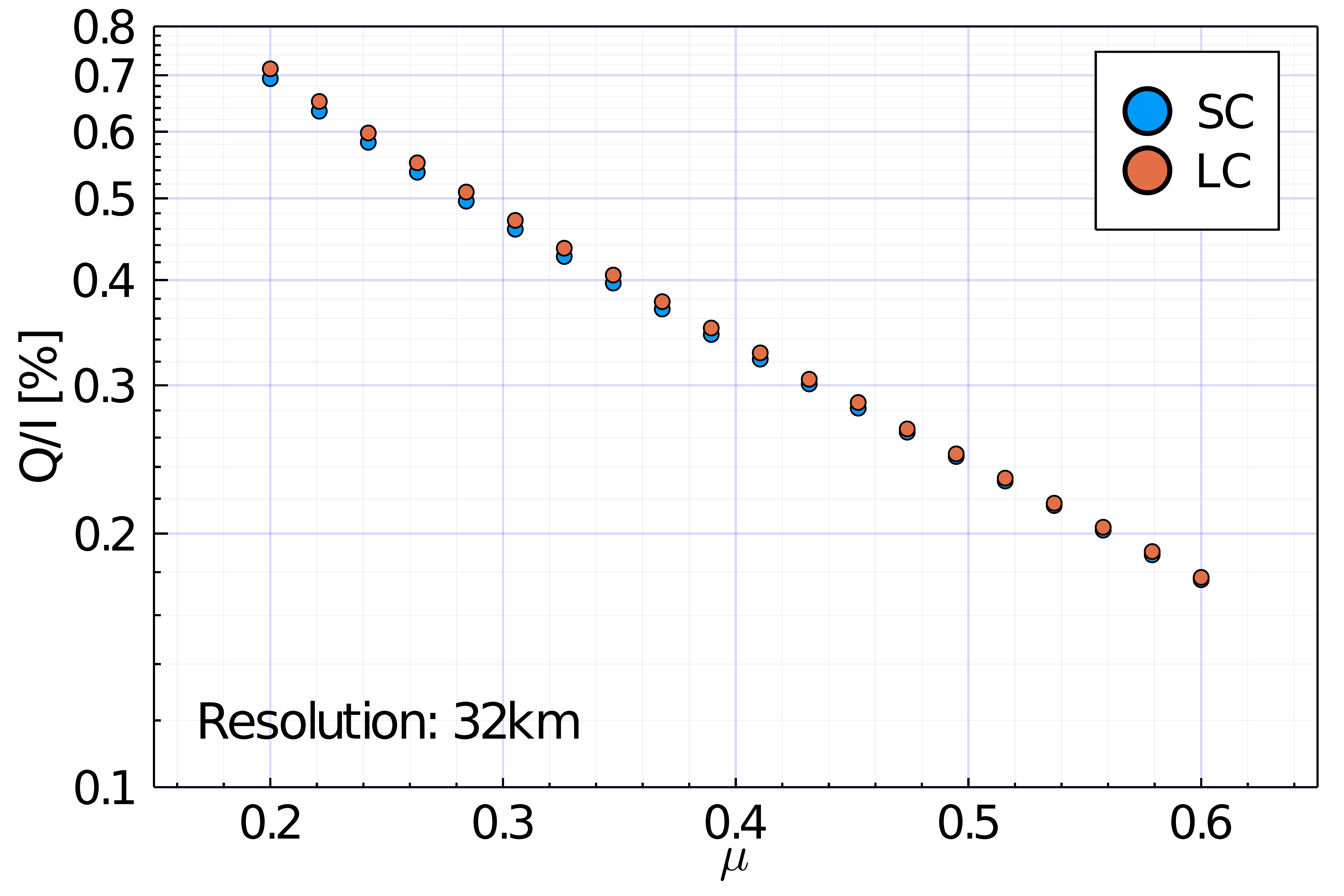}{0.48\textwidth}{(b)}}
\gridline{\fig{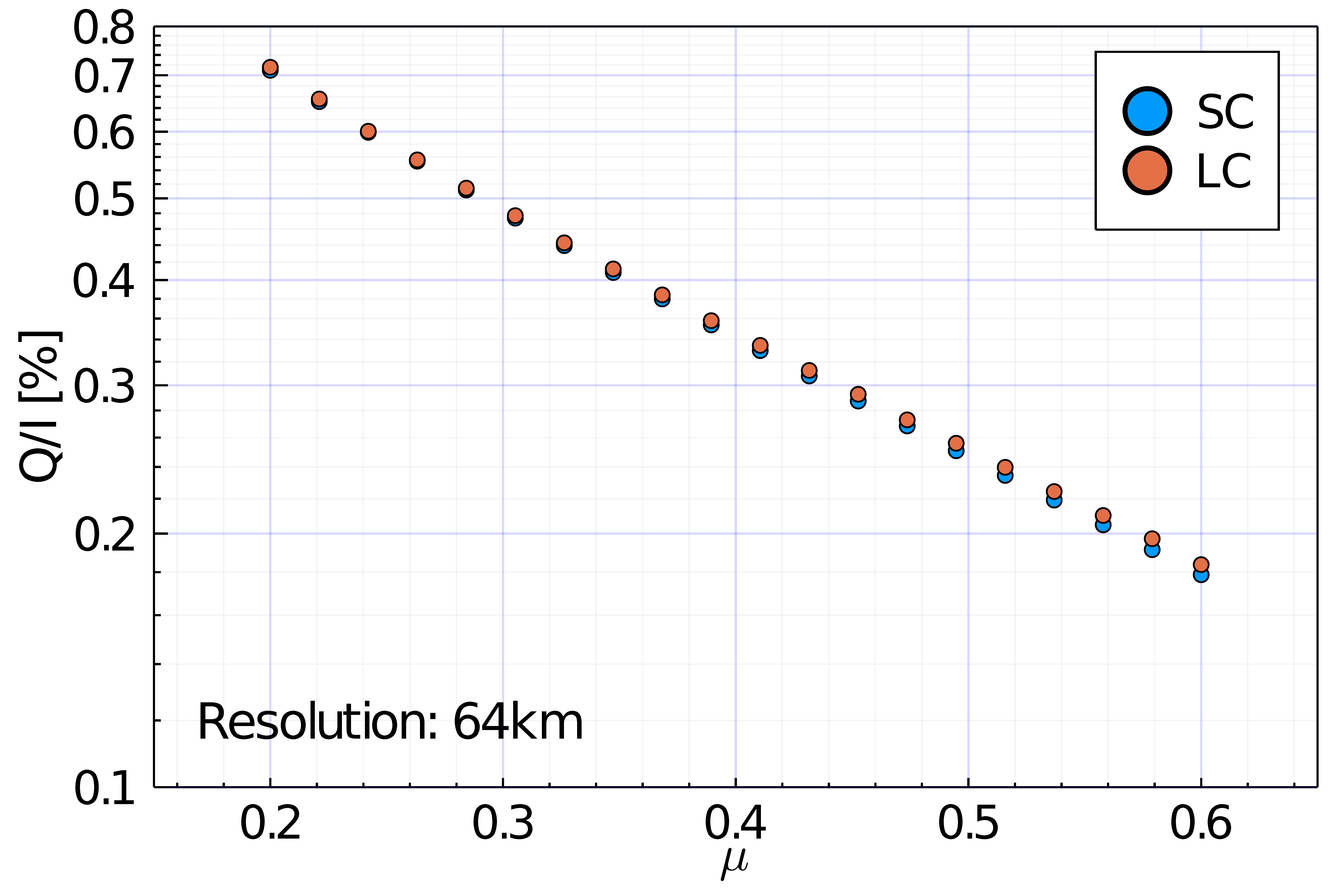}{0.48\textwidth}{(c)}
          \fig{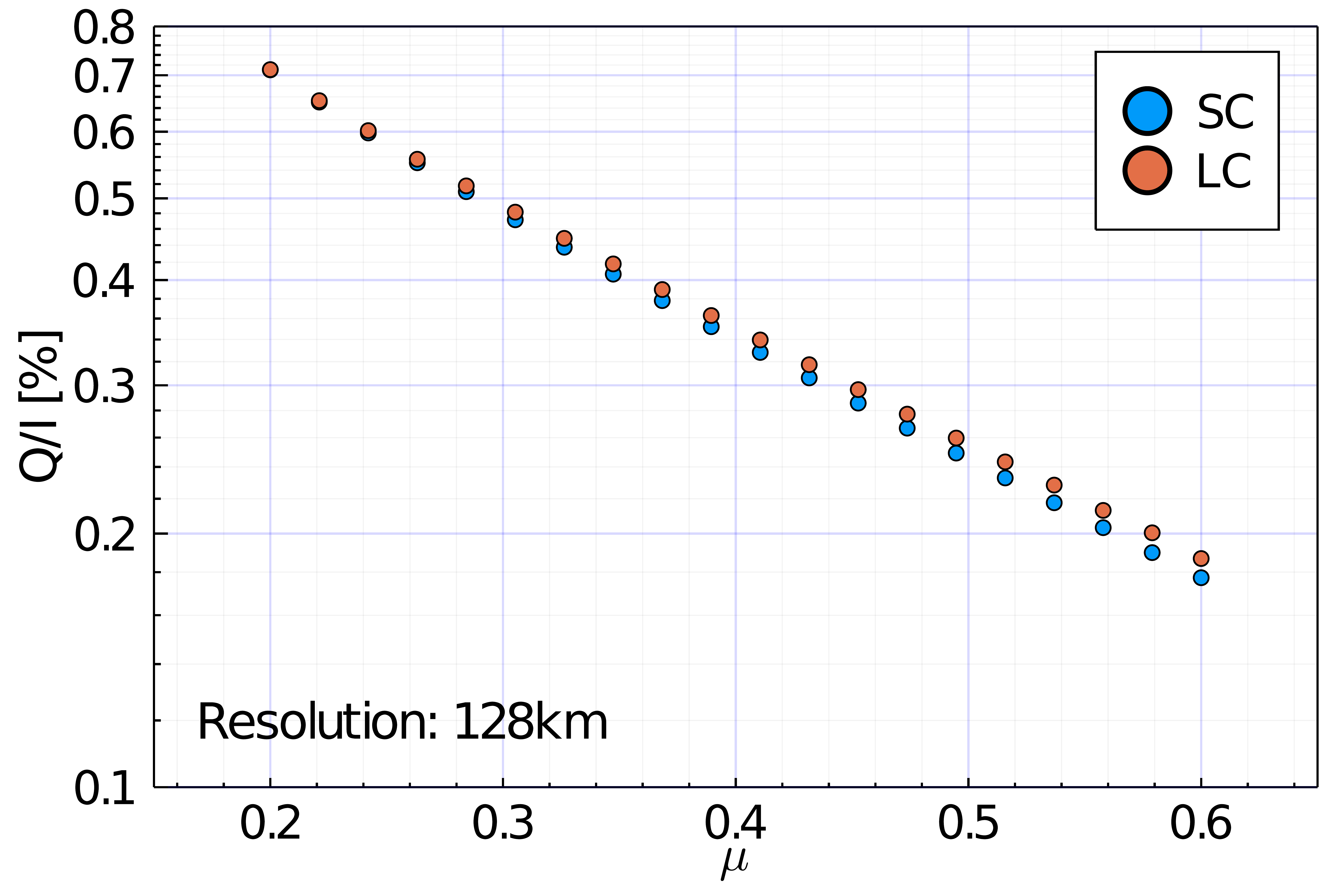}{0.48\textwidth}{(d)}}
\caption{The \gls*{clv} of the spatially integrated fractional linear
  polarization $Q/I$ obtained with the \gls*{sc} method (blue dots) and the \gls*{lc}
  method (orange dots) with a vertical axis resolution of $16$~km and a
  horizontal axis resolution of (a) $16$, (b) $32$, (c) $64$, and (d)
  $128$~km. Each data point in the figure was obtained by integrating
    over twelve equally spaced azimuths and over the whole field of view
    corresponding to each line of sight characterized by $\mu$.}
\label{fig:fig_clv_lc_vs_sc_lowres}
\end{figure*}

First, we study the impact on the \gls*{clv} of the linear polarization
\citep[\figa 12 in][]{DelPinoAleman2018}. Because the calculated \gls*{clv}
lacks spatial resolution (that is, for every line of sight the whole field of
view was integrated), the effects of the dispersion should be attenuated. This
is exactly what is found in \figa \ref{fig:fig_clv_lc_vs_sc}, which shows the
\gls*{clv} of the fractional linear polarization $Q/I$ computed in the $8$~km
grid using the \gls*{sc} and \gls*{lc} methods in the formal solver. For both Figs. \ref{fig:fig_clv_lc_vs_sc} and
  \ref{fig:fig_clv_lc_vs_sc_lowres},  we considered twenty equally spaced $\mu$
  values in the range [0.2,0.6] and, for each of them, twelve equally spaced
  azimuths. The differences between the two calculations are visible at plot level. However,
they are small enough so as to be irrelevant in the analysis, as the variations
due to the inhomogoneities of the solar atmosphere and the errors of the
observational data are larger than the difference in the results obtained with
such two formal solution methods.

It is of interest, however, to find out if this good agreement between the
\gls*{clv} calculated with the two methods is due to the high resolution of the
model atmosphere. To this end, we degrade the model by decreasing the resolution
of the vertical direction to $16$~km, and the resolution in the horizontal
directions to $16$, $32$, $64$, and $128$~km.
 This degradation was performed by resampling the model, taking every second point when
 going from $8$ to $16$~km resolution, and again (but only in the horizontal dimensions)
 from $16$ to $32$~km, from $32$ to $64$~km, and from $64$ to $128$~km resolutions.
 Interestingly, while the
comparison indicates that the worst resolution leads to larger differences
between the \gls*{sc} and \gls*{lc} results, they do not show a clear tendency
when the resolution is deteriorated (see \figa
\ref{fig:fig_clv_lc_vs_sc_lowres}). We can thus conclude that the \gls*{clv} of
the spatially integrated Stokes parameters is not strongly affected by the
dispersion effect of the \gls*{sc} algorithm.

\subsection{Impact on the spatially resolved polarization maps}

Secondly, we study the impact on the fractional total linear polarization maps
for an inclined \gls*{los}. In particular, we choose a \gls*{los} with
$\mu=0.5$, with $\mu$ the cosine of the heliocentric angle (see the panels of
the middle column in \figa 9 of \citealt{DelPinoAleman2018}, which they obtained
with the \gls*{sc} formal solver in the $8$~km resolution model).

\begin{figure*}
\gridline{\fig{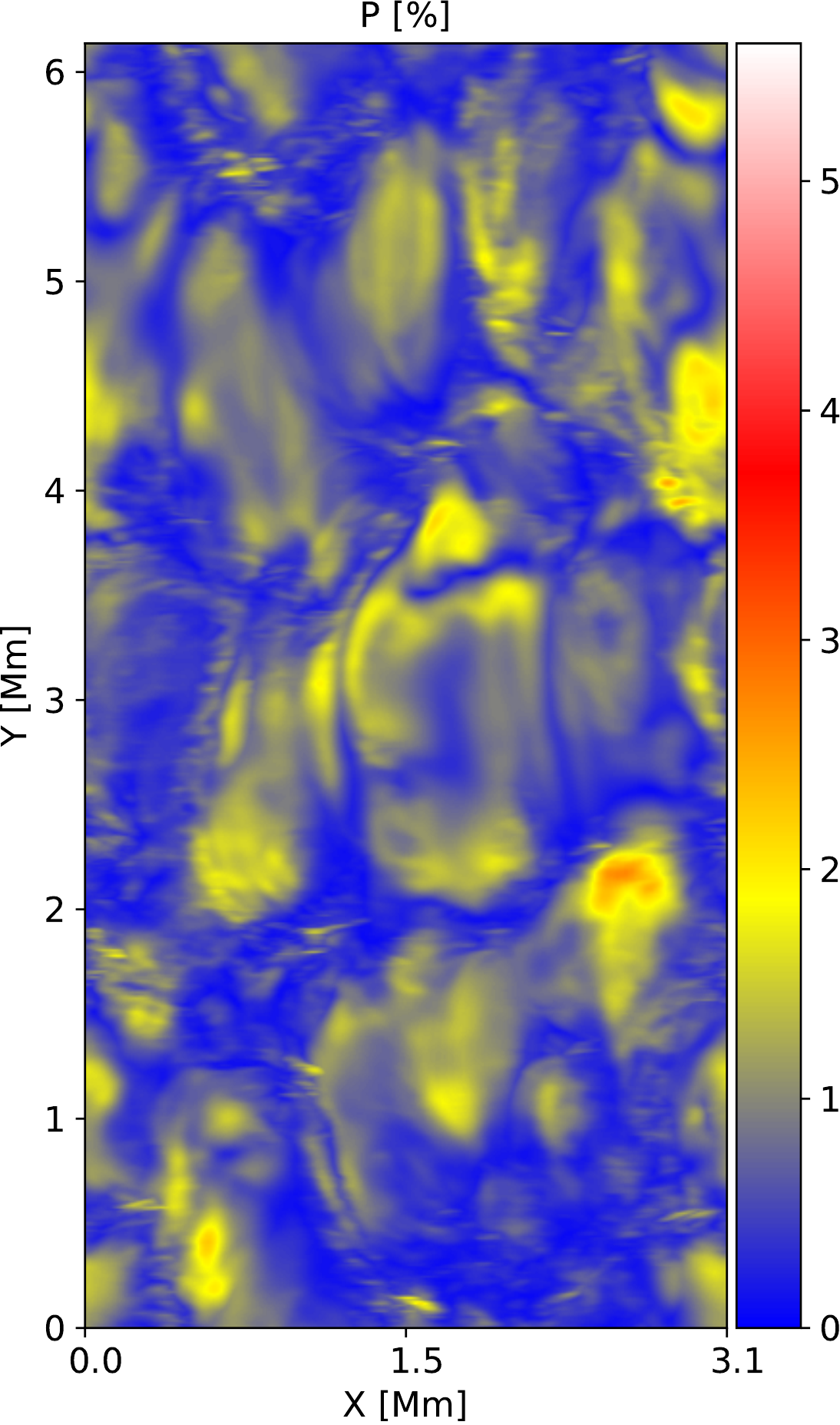}{0.38\textwidth}{(a)}
  \fig{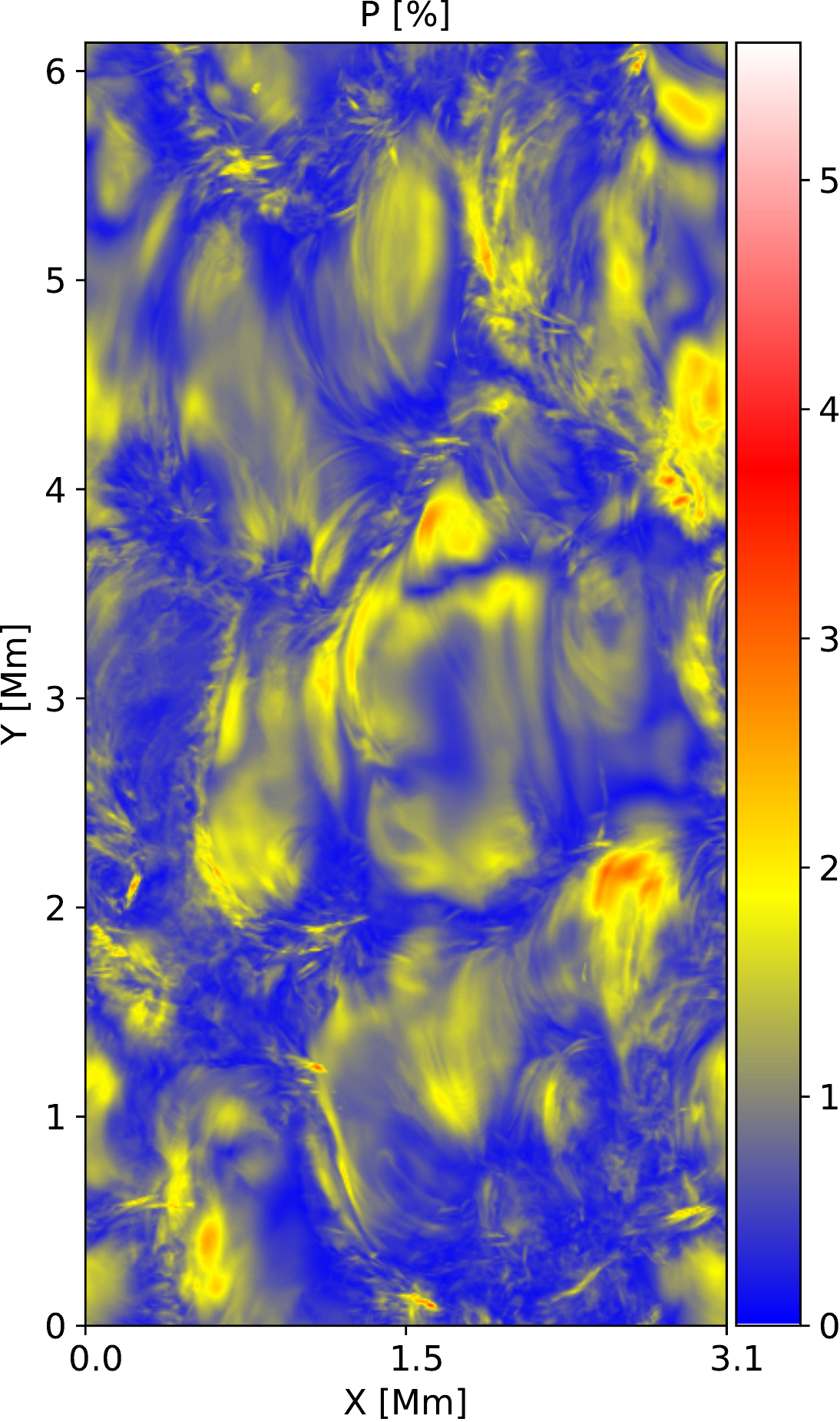}{0.38\textwidth}{(b)}}
\gridline{\fig{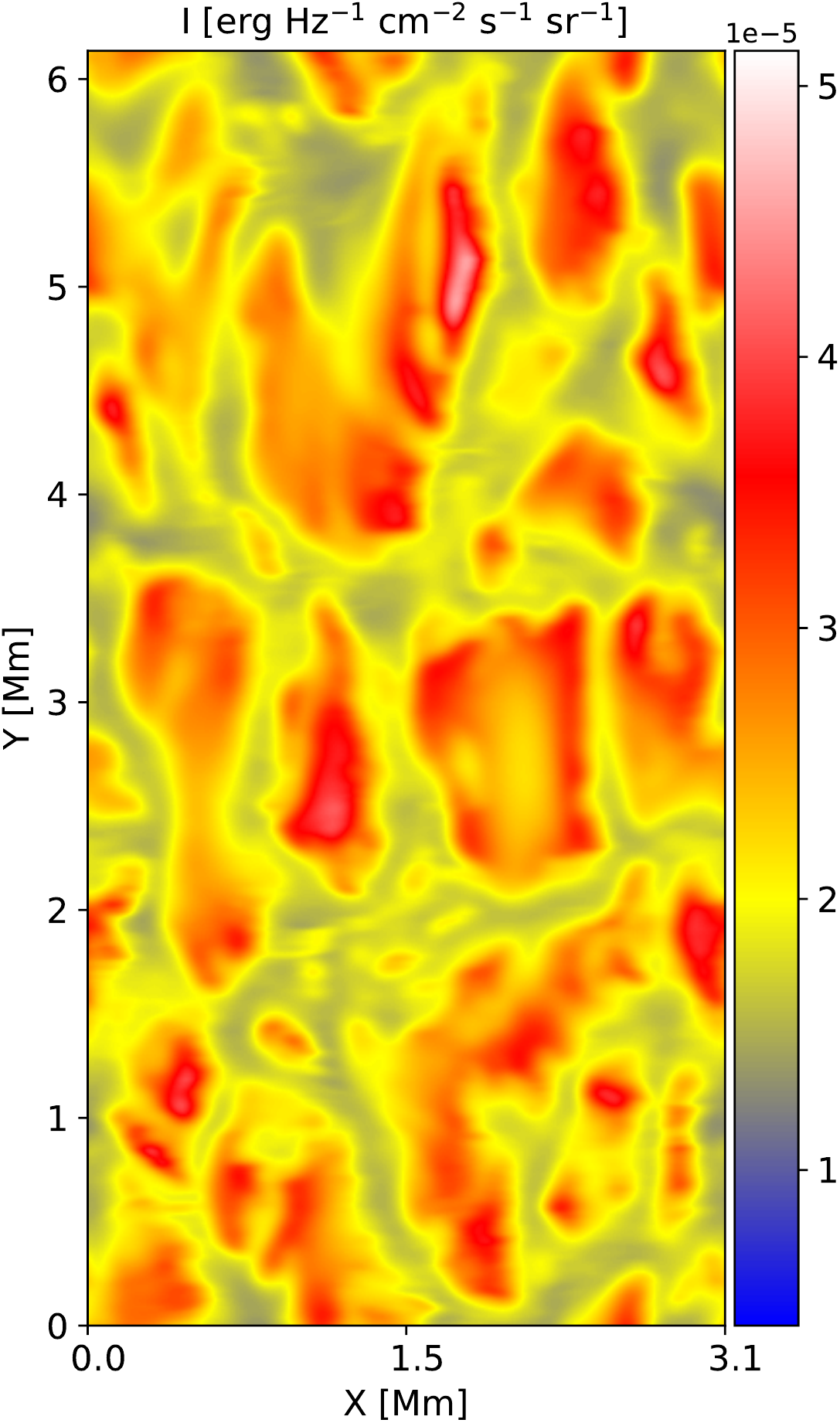}{0.38\textwidth}{(c)}
          \fig{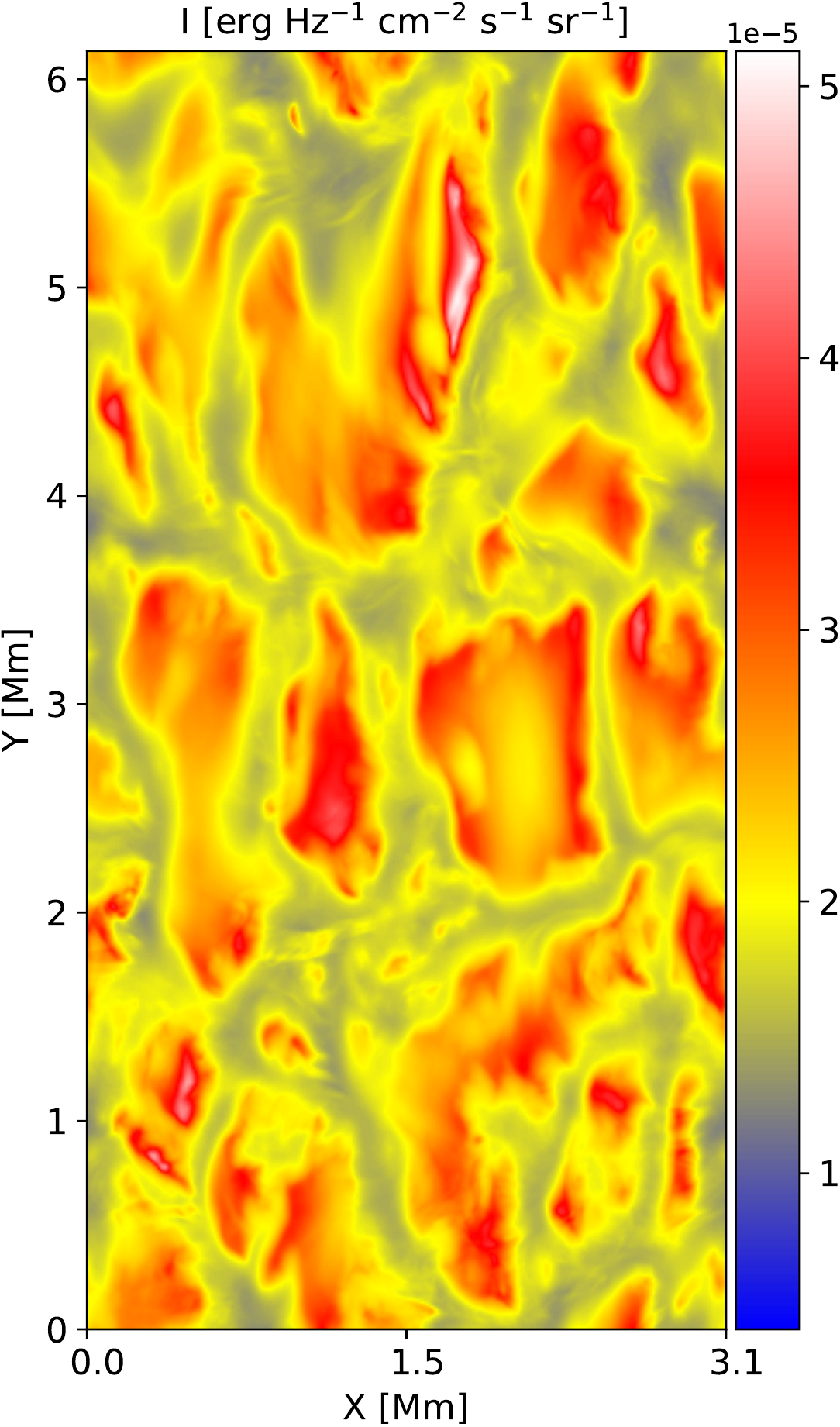}{0.38\textwidth}{(d)}}
\caption{Fractional total linear polarization amplitude (top panels) and continuum intensity amplitude (bottom panels) of the \ion{Sr}{1} 4607~\AA\ line
calculated for a LOS with $\mu=0.5$ in the full resolution ($8$~km) 3D model 
using the \gls*{sc} (panels a and c) and \gls*{lc} (panels b and d)
  formal solvers. Note
  that the $X$ coordinate in the figure represents the projected $X$ for the
  $\mu=0.5$ LOS, and not that of the original model.}
\label{fig:fig_sc_vs_lc_emergent}
\end{figure*}

\figa \ref{fig:fig_sc_vs_lc_emergent} shows (top panels) the fractional total linear polarization ($\sqrt{Q^2 + U^2}/I$) and the
continuum intensity amplitude (bottom panels) maps obtained with the \gls*{sc}
and the \gls*{lc} formal solvers. When using the \gls*{lc} solver (right panels),
the very fine and detailed structures produced by the model atmosphere are fully
preserved. However, when using the \gls*{sc} method (left panels) the numerical dispersion
introduces a clear ``smearing'' effect in the resulting polarization maps.

While this test indicates that, visually, the polarization of the emergent
radiation is sensitive to the the formal solver chosen, a more quantitative
study is necessary, which is addressed in the following section.

\subsection{Impact on the statistics}

Finally, we compare the impact on the scatter plots of linear polarization vs
continuum intensity after calculating the emergent Stokes profiles of the
\ion{Sr}{1} 4607~\AA\ line. To this end, we have selected the largest signals of
the $Q/I$ and $U/I$ profiles of the emergent spectral line radiation for each
spatial point of the disk-center field of view, and represented them against the
scaled value of the continuum intensity at the same point (see the top panels of
the \figa 18 of \citealt{DelPinoAleman2018}).

\begin{figure*}
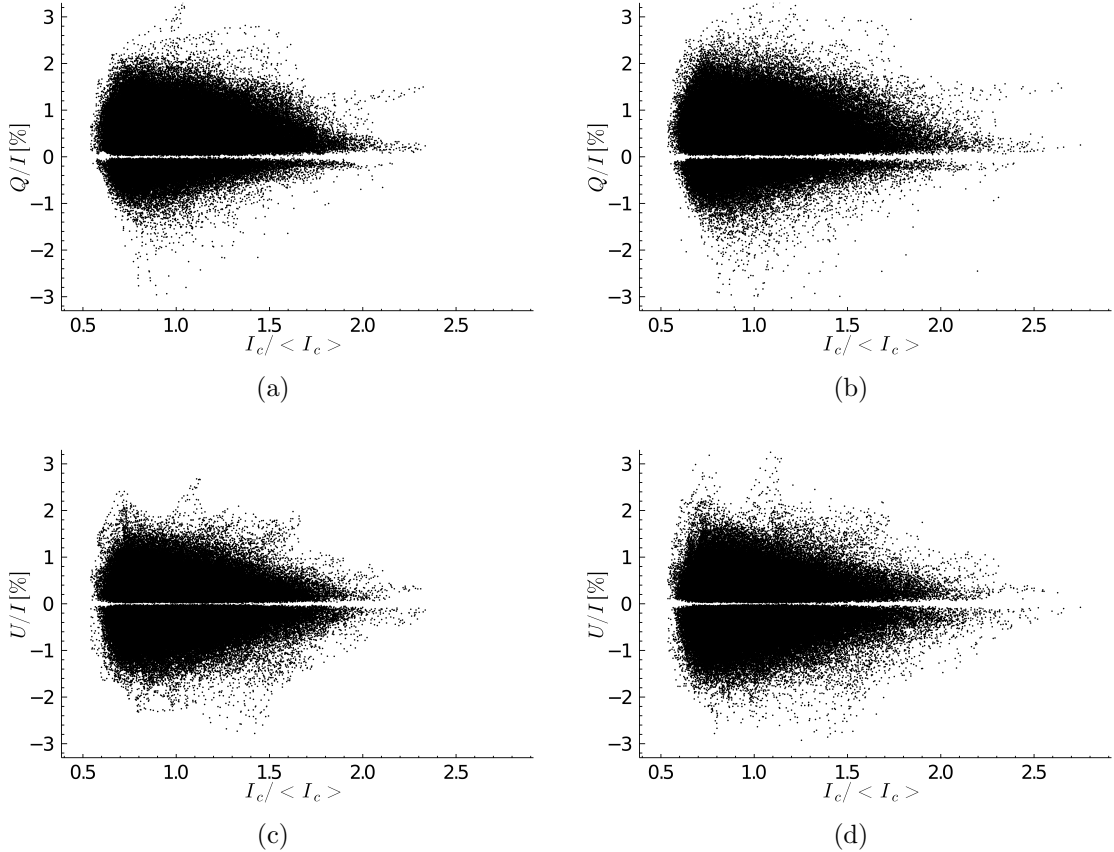

\gridline{\fig{mu0.5_km8_B_SC_max_Q.pdf}{0.48\textwidth}{(a)}
          \fig{mu0.5_km8_B_LC_max_Q.pdf}{0.48\textwidth}{(b)}}
\gridline{\fig{mu0.5_km8_B_SC_max_U.pdf}{0.48\textwidth}{(c)}
          \fig{mu0.5_km8_B_LC_max_U.pdf}{0.48\textwidth}{(d)}}
\caption{$Q/I$ (top panels) and $U/I$ (bottom panels) amplitudes of the emergent
  radiation against the scaled value of the continuum intensity calculated with
  the \gls*{sc} (panels a and c) and \gls*{lc} (panels b and d) formal
  solvers. The line of sights have $\mu=0.5$ and eight equally spaced
  azimuths. The resolution of the model's grid is $8$~km. }
\label{fig:fig18-8-B}
\end{figure*}

\figa \ref{fig:fig18-8-B} shows these scatter plots for the original resolution
($8$~km) of the atmospheric model and $\mu=0.5$. While the selection of
algorithm for the formal solver does not impact the behaviour (the structure) of
the distribution of the points, there are some differences.  First, ``trails''
are more frequent with \gls*{sc}. These trails seem to appear as a consequence
of the foreshortening of one of the dimensions for inclined \gls*{los}. Along
the shortened direction, nodes are closer to each other in the field of view
and, therefore, the change of values between nodes is larger in one dimension
than in the other, giving rise to the ``trails'' in the scatter plots.

Secondly, the maximum values of the polarization are larger with \gls*{lc}.
This is a consequence of the smearing effect due to the dispersion introduced by
the \gls*{sc} algorithm, which smudges the regions with those maximums and
decreases its amplitudes.

\begin{figure*}
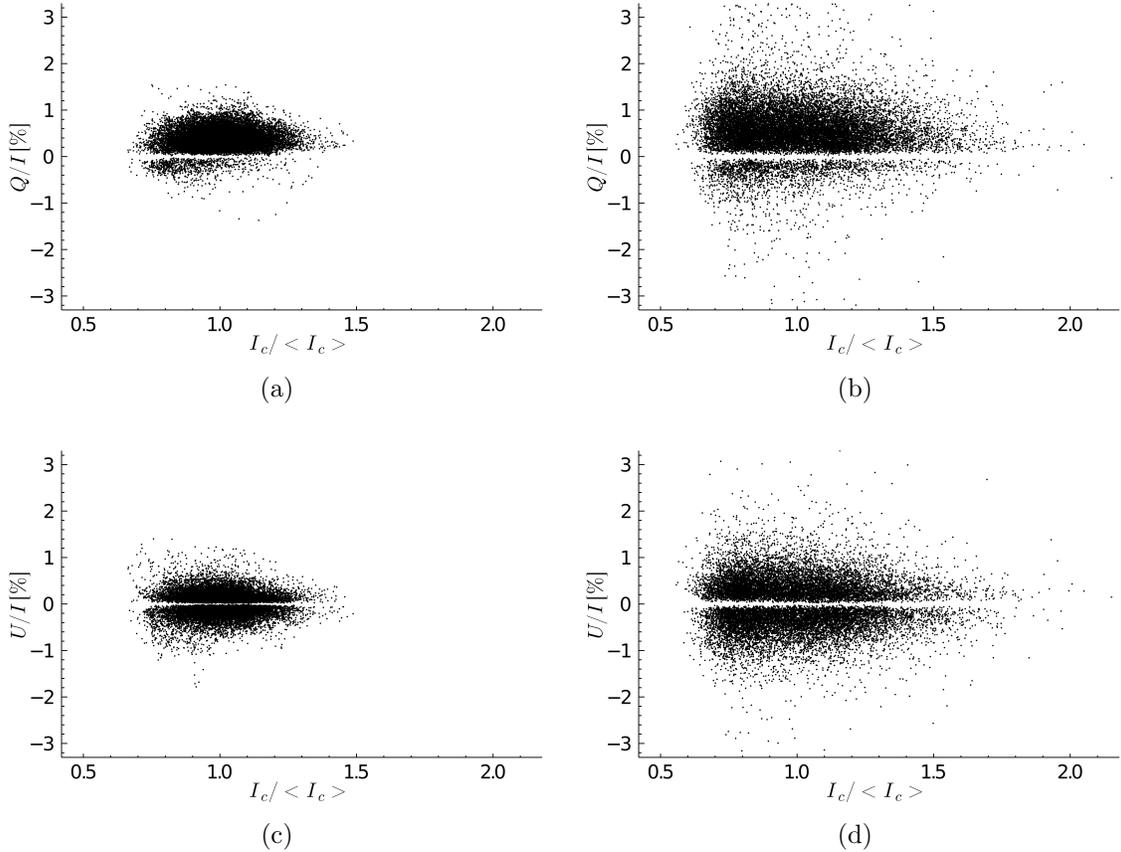

\gridline{\fig{mu0.5_km128_dz16_B_SC_max_Q.pdf}{0.48\textwidth}{(a)}
          \fig{mu0.5_km128_dz16_B_LC_max_Q.pdf}{0.48\textwidth}{(b)}}
\gridline{\fig{mu0.5_km128_dz16_B_SC_max_U.pdf}{0.48\textwidth}{(c)}
          \fig{mu0.5_km128_dz16_B_LC_max_U.pdf}{0.48\textwidth}{(d)}}
\caption{$Q/I$ (top panels) and $U/I$ (bottom panels) amplitudes of the emergent
  radiation against the scaled value of the continuum intensity calculated with
  the \gls*{sc} (panels a and c) and \gls*{lc} (panels b and d) formal solvers.
  The line of sights have $\mu=0.5$ and eight equally spaced azimuths.  The
  resolution of the grid is $128$~km along the horizontal directions and $16$~km
  along the vertical.}
\label{fig:fig18-128-B}
\end{figure*}

In \figa \ref{fig:fig18-128-B} we can see the case with horizontal resolution of
$128$~km and vertical resolution of $16$~km.  For this resolution we see that
the differences are larger than in the $8$~km resolution case, changing not only
the maximum signals but also altering the shape of the distribution of
points. In the low resolution case the maximum values of the polarization
signals are larger with \gls*{lc}, and the SC results do not show so clearly the
inverse relation between the polarization amplitude and the continuum intensity
that \citealt{DelPinoAleman2018} found in their $8$~km resolution calculations.

Regarding the total linear polarization (see \figa 10 of
\citealt{DelPinoAleman2018}), \figa \ref{fig:P-8-B} shows such scatter plots in
the full resolution ($8$~km) 3D model applying the \gls*{sc} (top panels) and
\gls*{lc} (bottom panels) formal solvers.  Three lines of sight with $\mu=0.1$
(left panels), $\mu=0.5$ (middle panels), and $\mu=1$ (right panels) are
shown. Similarly to what we observe in \figa \ref{fig:fig18-8-B}, the main
difference between the \gls*{sc} (top panels) and \gls*{lc} (bottom panels)
solutions is a slight reduction of the maximum polarization values in the
\gls*{sc} solution, but the anti-correlation between the total linear
polarization amplitude and the continuum intensity pointed out by
\citet{DelPinoAleman2018} is confirmed by the \gls*{lc} results. Obviously, the
results for $\mu=1$ are identical because, as explained in section
\S\ref{sec:vacuum}, when the propagation direction is along the vertical all
points along the characteristics correspond to nodes and thus there is neither
dispersion nor differences between the \gls*{sc} and \gls*{lc} solutions.

\begin{figure*}
  \gridline{\fig{mu0.1_km8_B_SC_max_P.pdf}{0.33\textwidth}{(a)}
    \fig{mu0.5_km8_B_SC_max_P.pdf}{0.33\textwidth}{(b)}
    \fig{mu1_km8_B_SC_max_P.pdf}{0.33\textwidth}{(c)}}
  \gridline{\fig{mu0.1_km8_B_LC_max_P.pdf}{0.33\textwidth}{(d)}
    \fig{mu0.5_km8_B_LC_max_P.pdf}{0.33\textwidth}{(e)}
    \fig{mu1_km8_B_LC_max_P.pdf}{0.33\textwidth}{(f)}}
\caption{Fractional total linear polarization $P$ (at the wavelength where $P$
  is maximum) against the scaled value of the continuum intensity calculated
  with the \gls*{sc} (top panels) and \gls*{lc} (bottom panels) formal
  solvers. The line of sights have $\mu=0.1$ (left panels), $\mu=0.5$ (middle
  panels) and $\mu=1$ (right panels). Eight equally spaced azimuths were taken
  for the $\mu=0.1$ and $\mu=0.5$ cases.  The resolution of the model's grid is
  $8$~km.}
\label{fig:P-8-B}
\end{figure*}

\begin{figure*}
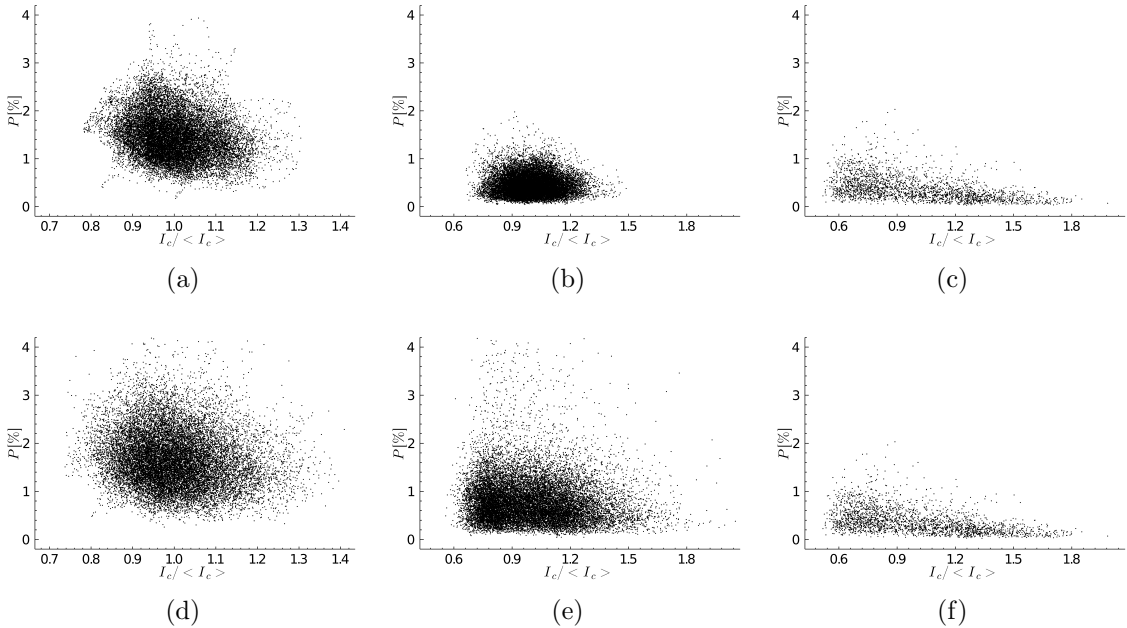

  \gridline{\fig{mu0.1_km128_dz16_B_SC_max_P.pdf}{0.33\textwidth}{(a)}
    \fig{mu0.5_km128_dz16_B_SC_max_P.pdf}{0.33\textwidth}{(b)}
    \fig{mu1_km128_dz16_B_SC_max_P.pdf}{0.33\textwidth}{(c)}}
  \gridline{\fig{mu0.1_km128_dz16_B_LC_max_P.pdf}{0.33\textwidth}{(d)}
    \fig{mu0.5_km128_dz16_B_LC_max_P.pdf}{0.33\textwidth}{(e)}
    \fig{mu1_km128_dz16_B_LC_max_P.pdf}{0.33\textwidth}{(f)}}
\caption{Fractional total linear polarization $P$ (at the wavelength where $P$
  is maximum) against the scaled value of the continuum intensity calculated
  with the \gls*{sc} (top panels) and \gls*{lc} (bottom panels) formal
  solvers. The line of sights have $\mu=0.1$ (left panels), $\mu=0.5$ (middle
  panels) and $\mu=1$ (right panels). Eight equally spaced azimuths were taken
  for the $\mu=0.1$ and $\mu=0.5$ cases.  The resolution of the grid is $128$~km
  along the horizontal directions and $16$~km along the vertical.}
\label{fig:P-128-B}
\end{figure*}

In \figa \ref{fig:P-128-B} we show the case with horizontal resolution of
$128$~km and vertical resolution of $16$~km.  For this resolution we can see,
similarly to \figa \ref{fig:fig18-128-B}, that the differences are larger than
in the $8$~km resolution case. The maximum values of the polarization are larger
with \gls*{lc}, except for $\mu=1$, which are identical in both solutions (see
explanation above).

\section{Conclusions}\label{sec:conclusions}

Although in 3D models of the solar atmosphere it is unfeasible (due to its
scaling) to solve the full iterative problem of scattering line polarization
using the \gls*{lc} instead of the \gls*{sc} method, the computational cost
increase for a single formal solution is not that significant. We have thus
implemented the \gls*{lc} formal solution method for the calculation of the
emergent Stokes parameters in the public radiative transfer code PORTA. Starting
from the self-consistent solution for the atomic excitation in the \ion{Sr}{1}
4607~\AA\ line calculated by \cite{DelPinoAleman2018} in a realistic 3D
magneto-convection model, we have computed and compared the emergent Stokes
profiles of this spectral line using both the \gls*{sc} and \gls*{lc}
algorithms.

When comparing polarization quantities without spatial resolution, such as the
\gls*{clv}, we have found that, as expected, there is little to no impact on the
selection of formal solvers, because the lack of spatial resolution conceals the
smearing effect of the \gls*{sc} method.

The impact is however noticeable when comparing both results at full spatial
resolution. The \gls*{lc} preserves the fine structure of the polarization maps
produced by the inhomogeneities in the \gls*{3d} model atmosphere, and it
generally yields larger polarization signals. In contrast, the \gls*{sc} method
damps and smears the polarization signals of the emergent spectral line
radiation, because of the numerical diffusion inherent to this method.

A comparison of the statistics (scatterplots of polarization against the
continuum intensity) also reveals that the main difference between the \gls*{sc}
and \gls*{lc} solutions, at the original resolution of the model, is just a
slight reduction of the maximum values of the scattering polarization signals,
keeping the overall shape of the distribution unchanged. However, the
differences increase when the spatial resolution of the model is deteriorated,
changing not only the maximum signals but also altering the shape of the
distribution of points.

We can thus conclude that, due to the high resolution of the \gls*{3d} model
atmosphere used in \cite{DelPinoAleman2018}, the results shown in
that paper
regarding the comparison with the observed \gls*{clv} or the statistical
relation between polarization signals and continuum intensities are not affected
by the use of the \gls*{sc} method. Nevertheless, we recommend the use of the
\gls*{lc} formal solver to compute the emergent Stokes profiles, avoiding the
smearing effect intrinsic to the \gls*{sc} method without a significant increase
on computational cost.

\acknowledgements

We acknowledge the funding received from the European Research Council (ERC)
under the European Union's Horizon 2020 research and innovation programme (ERC
Advanced Grant agreement No 742265), as well as through the project
PGC2018-095832-B-I00 of the Spanish Ministry of Science, Innovation and
Universities. The 3D radiative transfer simulations were carried out with the
MareNostrum supercomputer of the Barcelona Supercomputing Center (National
Supercomputing Center, Barcelona, Spain), and we gratefully acknowledge the
technical expertise and assistance provided by the Spanish Supercomputing
Network, as well as the additional computer resources used, namely the La Palma
Supercomputer located at the Instituto de Astrofísica de Canarias.


\bibliography{lc}{}
\bibliographystyle{aasjournal}

\end{document}